\RequirePackage{snapshot}
\documentclass[
reprint,
superscriptaddress,
 amsmath,amssymb,
 aps,longbibliography,floatfix
]{revtex4-2}

\usepackage{graphicx}
\usepackage{dcolumn}
\usepackage{bm}
\usepackage{color,amsmath}
\usepackage{physics}
\usepackage{subfiles}
\usepackage{bm}
\usepackage{siunitx}
\usepackage{enumitem}

\usepackage[english]{babel}

\usepackage[normalem]{ulem}
\usepackage[usenames,dvipsnames]{xcolor}
\usepackage{hyperref}
\hypersetup{
    colorlinks,
    linkcolor={violet},
    citecolor={ForestGreen},
    urlcolor={teal}
}

\usepackage{subfigure}

\newcommand{\perceptrain}{PN }
\newcommand{\perceptrainn}{PN}

\begin{document}

\title{Hybrid between biologically and quantum-inspired many-body states}

\author{Miha Srdinšek}
\email{mihasrdinsek@gmail.com}
\affiliation{Universit\'e Grenoble Alpes, CEA, Grenoble INP, IRIG, Pheliqs, F-38000 Grenoble, France}
\author{Xavier Waintal}
\affiliation{Universit\'e Grenoble Alpes, CEA, Grenoble INP, IRIG, Pheliqs, F-38000 Grenoble, France}
\date{\today}

\begin{abstract}
Deep neural networks can represent very different sorts of functions, including complex quantum many-body states. Tensor networks can also represent these states, have more structure and are easier to optimize. However, they can be prohibitively costly computationally in two or higher dimensions. Here, we propose a generalization of the perceptron -- the perceptrain -- which borrows features from the two different formalisms. We construct variational many-body ansatz from a simple network of perceptrains. The network can be thought of as a neural network with a few distinct features inherited from tensor networks. These include efficient local optimization akin to the density matrix renormalization algorithm, instead of optimizing all the parameters at once; the possibility to dynamically increase the number of parameters during the optimization; the possibility to compress the state; and a structure that remains quantum-inspired. We showcase the ansatz using a combination of variational Monte Carlo (VMC) and Green function Monte Carlo (GFMC) on a $10\times 10$ transverse field quantum Ising model with a long-range $1/r^6$ antiferromagnetic interaction. The model corresponds to the Rydberg (cold) atoms platform proposed for quantum annealing. We consistently find a very high relative accuracy for the ground state energy, around $10^{-5}$ for VMC and  $10^{-6}$ for GFMC in all regimes of parameters, including in the vicinity of the quantum phase transition. We use very small ranks ($\sim 2$-$5$) of perceptrains, as opposed to multiples of thousand used in matrix product states. The optimization of the energy was very robust. The entire phase diagram was found with a single initial condition and a fixed set of hyperparameters.
\end{abstract}

\maketitle

\section{Introduction}
\label{sec:Introduction}

Two different sorts of networks are profoundly reshaping the landscape of numerical computations. On one hand, neural networks (biologically inspired) lie at the core of artificial intelligence \cite{vaswani_attention_2023}. On the other hand, tensor networks (quantum inspired), a natural framework for describing complex quantum states, are being increasingly applied in unrelated fields ranging from plasma physics to turbulence and optimization \cite{nunez_fernandez_learning_2025}. Very roughly speaking, neural networks promise generality and an algorithmic paradigm shift, since a large part of the learning process is still not understood. They seem to have a boundless range of applications. Instead, tensor networks promise access to exponentially large objects in a highly structured way. They are built on a well understood data compression, based on entanglement. Tensor networks are perhaps less general than neural networks, but when they can be applied, their internal structure makes them much more powerful. The goal of this article is to explore hybrid networks that retain \emph{some} of the interesting properties of tensor networks in exchange for \emph{some} of the added generality of the neural networks.

More specifically, both types of networks have been highly successful as variational ans\"atze for describing the ground states of quantum many-body systems.

For one-dimensional (1D) systems, matrix product states (MPS), also known as tensor trains, together with the associated Density Matrix Renormalization Group algorithm (DMRG) \cite{White1992a,White1992b} have allowed one to essentially solve a number of problems with machine precision. MPS are the most studied tensor networks. Their extension to two dimensional (2D) systems, projected entangled-pair states (PEPS), are theoretically very promising but their practical application is hindered by a large computational cost \cite{corboz_simulation_2010}, so that two dimensional systems remain in practice very challenging. One could say that these techniques are on the verge of being able to tackle 2D correlated systems \cite{xu_coexistence_2024,qian_clifford_2024}, a central goal of contemporary computational many-body physics.

Using neural networks to represent many-body states, the so-called neural network quantum states (NNQS), is much more recent, yet quite significant successes have been demonstrated \cite{carleo_solving_2017,nomura_restricted_2017,vicentini_variational_2019,choo_two-dimensional_2019,hibat-allah_recurrent_2020,lovato_hidden-nucleons_2022,robledo_moreno_fermionic_2022,sharir_neural_2022,kim_neural-network_2023}. These ans\"atze are very general. The absence of tensor network structure means that one has to resort to Variational Monte Carlo (VMC) to sample the wave function and estimate its energy and gradient with respect to the ansatz parameters. The ground state is found through some variant of (noisy) gradient descent. Although these ans\"atze are believed to be general enough to capture the ground state accurately, the challenge is in the optimization which can become a very challenging task \cite{chen_empowering_2024}. This is a consequence of the non-physicality, a.k.a. over-generality, of the ansatz, with up to hundreds of thousands of parameters. Also, the structure of the ansatz is typically fixed at the beginning of the calculation, and it is not straightforward to dynamically change the number of parameters of the ansatz \cite{dash_efficiency_2025}, a feature at the root of the success of DMRG. 

The hypothesis at the root of the present work is that variational ans\"atze used in VMC follow some sort of "expressivity-trainability" duality: the more general ("expressivity") the ansatz is, in terms of its (lack of) structure and bare number of parameters, the more likely it will be able to approach an arbitrary ground state, but at the cost of being more difficult to optimize ("trainability"). A precise version of this duality remains to be formulated, and its validity assessed. For the moment we should treat it simply as a guiding principle in the upcoming ansatz construction: there is probably some room between the rigid structure of MPS and the highly general neural networks such as transformers which are expressive enough to represent both a many-body ground state \emph{and} a large language model. To try and address this interplay, we build what we shall call a "perceptrain", a sort of perceptron that uses internally a tensor train. Then we make a simple many-body ansatz by building a simple neural network of perceptrains. This work bears some similarities with other recent VMC calculations including tensor networks ans\"atze such as PEPS \cite{sandvik_variational_2007,liu_accurate_2021,naumann_rizzi_introduction_2024}, tensorial neural networks (TNN) \cite{wang_tensor_2023,wu_tensor-network_2023}, matrix product backflow states (MPBS) \cite{lami_matrix_2022}, MPS \cite{sandvik_variational_2007}, and also with the so-called string bond states (SBS) ansatz \cite{schuch_strings_2008,sfondrini_simulating_2010,glasser_neural-network_2018}.
 
The paper is structured as follows. In Sec.~\ref{sec:Perceptrain} we introduce the perceptrain and comment on the new features it brings to a network. In Sec.~\ref{subsec:Perceptrain-for-spin-models} we discuss how several perceptrains can be used to construct many-body wave functions. We propose a tailored optimization path for these ans\"atze, reminiscent of the DMRG algorithm. In Sec.~\ref{sec:Application-to-Spin-models} we use the ansatz to solve a $10\times 10$ transverse field quantum Ising model. In Sec.~\ref{sec:technical} we explain more technical details about spontaneous symmetry breaking in VMC, our GFMC benchmark, and compare different optimization approaches. Finally, in Sec.~\ref{sec:Conclusions} we share our concluding remarks.

\section{From perceptron to perceptrain}
\label{sec:Perceptrain}

\subsection{Notations for perceptrons and MPS.}

The basic building block of deep neural networks is the perceptron which mimics a biological neuron that "fires" when its depolarization potential reaches a certain threshold. The perceptron is a non-linear function $\phi(\boldsymbol{x})$ of $n$ real numbers $\boldsymbol{x}=(x_1, x_2, \dots, x_n)$, the inputs, 
\begin{equation}\label{eq:perceptron}
\phi(\boldsymbol{x}) = f(\boldsymbol{W} \cdot \boldsymbol{x}+b),
\end{equation}
where $\boldsymbol{W}$ is a vector of parameters (\emph{weights}) and the additional parameter $b$, the threshold (\emph{bias}). The function $f(x)$ is typically the Fermi function (a common alternative being ReLU)  mirrored along the y-axis (known as the logistic or sigmoid function in this context). Neural networks are constructed by chaining many network units (perceptrons), i.e. the output of one is used as the input of another.

The MPS is also a function of $n$ inputs that returns a scalar. However the input here is a vector $\boldsymbol{s}=(s_1, s_2, \dots, s_n)$ of indices that take \emph{discrete} values: $s_i \in \{1,2,...,d\}$. The "parameters" of the MPS are a set of $nd$ matrices $A_i^{s_i}$ and the MPS evaluates to,
\begin{equation}\label{eq:MPS}
\phi_{\boldsymbol{s}} =  \mathbf{Tr}  A_1^{s_1} A_2^{s_2}...A_n^{s_n}.
\end{equation}
The size of the matrix $A_i^{s_i}$ is $\chi_i \times \chi_{i+1}$ where $\chi_i$ is known as the bond dimension. By convention, $\chi_{1}=\chi_{n}=1$ so the first and last matrices are in fact vectors. The schematic of MPS is shown in Fig.~\ref{fig:drawing-perceptrain}b.

\subsection{Perceptrain construction.}

The perceptrain corresponds to replacing the inner function $\boldsymbol{W} \cdot \boldsymbol{x}$ of the perceptron with an MPS. To fully replace perceptrons, we define a generalization of the MPS, which may take continuous values as an input.

\begin{figure}[t]
    \centering
        \includegraphics[width = 1.0\linewidth]{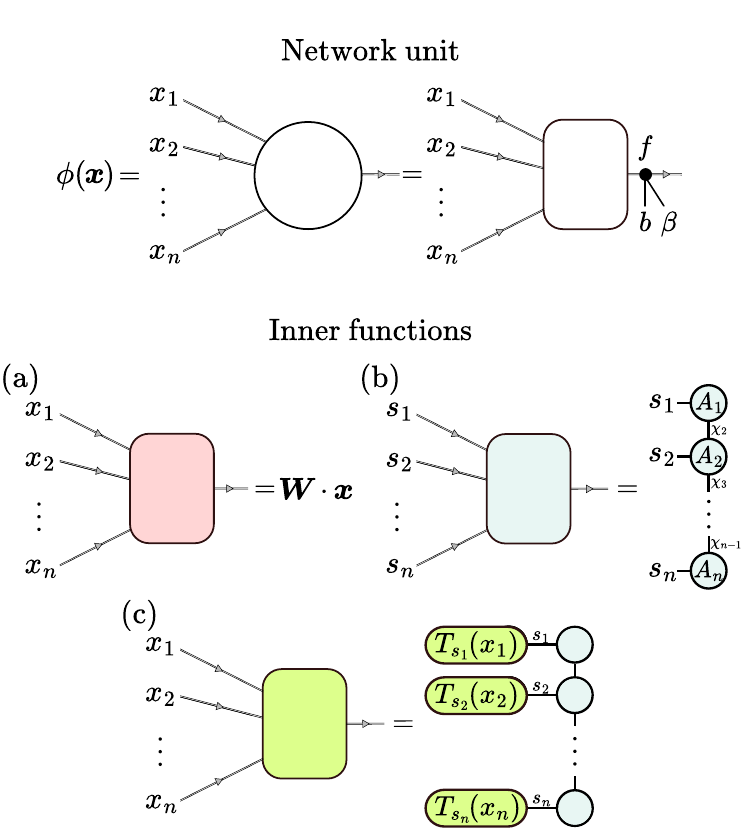}
    \caption{(Top) A network unit $\phi(\boldsymbol{x})$ (circular shape) is the combination of an activation function $f$ [e.g. $\tanh(\beta x +b)$, black dot] with an inner function (rounded rectangular shape). (Bottom) Different inner functions.
 (a) Inner function of the perceptron as in  Eq.~\eqref{eq:perceptron}, not used in this work.
 (b) MPS inner function as in Eq.~\eqref{eq:MPS} for discrete inputs.
 (c) Continuous MPS inner function as in Eq.~\eqref{eq:perceptrain1}. 
 Note that in the left- and right- hand side of panels (b) and (c), the lines have different meaning; see text.
    \label{fig:drawing-perceptrain}}
\end{figure}

The generalization can be done in many ways. Here we choose to expand the perceptrain on a basis of Chebyshev polynomials $T_s(x)$ of the first kind. Since these polynomials are well behaved for inputs inside the interval $x\in [-1,+1]$, we use the hyperbolic tangent to convert the output of perceptrains to this interval. The overall network unit function in the continuum reads,
\begin{equation}\label{eq:perceptrain1}
\phi(\boldsymbol{x}) = \mathbf{tanh} \left[b +\beta \sum_{\boldsymbol{s}} \phi_{\boldsymbol{s}}T_{s_1}(x_1)T_{s_2}(x_2)...T_{s_n}(x_n)\right]
\end{equation}
where $\phi_{\boldsymbol{s}}$ is an MPS and $\beta$ and $b$ are two additional parameters. Introducing the continuous matrix function $A_i(x)$ as
\begin{equation}\label{eq:perceptrain-sum}
A_i(x) = \sum_{s} A_i^s T_s(x),
\end{equation}
the perceptrain can also be rewritten as
\begin{equation}\label{eq:perceptrain}
\phi(\boldsymbol{x}) = \mathbf{tanh} \left[b +\beta \mathbf{Tr}  A_1(x_1)A_2(x_2)...A_n(x_n)\right].
\end{equation}
The upper panel of Fig.~\ref{fig:drawing-perceptrain} shows a schematic construction. A network unit (circular shape), is composed of an \emph{inner function} (rounded rectangular shape) and an activation function ($f$, or $\tanh$, black dot). We distinguish three types of inner functions: the one of the perceptron $\boldsymbol{W} \cdot \boldsymbol{x}$ [red, Figs.~\ref{fig:drawing-perceptrain}(a)], an MPS with discrete input values [blue, Figs.~\ref{fig:drawing-perceptrain}(b)] and an MPS with continuous input values (green, Figs.~\ref{fig:drawing-perceptrain}(c)). Note the \emph{different meanings} associated with the left hand side and the right hand side of Figs.~\ref{fig:drawing-perceptrain}(b) and \ref{fig:drawing-perceptrain}(c). The lines with arrows on the left hand side correspond to flows of values, while the legs of tensors represents their indices. If two tensors are connected with a leg (a line without an arrow), this represents a summation over the corresponding index. We refer to any network unit that contains a tensor train as the inner function, i.e. either Fig.~\ref{fig:drawing-perceptrain}(b) or Fig.~\ref{fig:drawing-perceptrain}(c), as a perceptrain.

\begin{figure}[t]
    \centering
        \includegraphics[width = 1.0\linewidth]{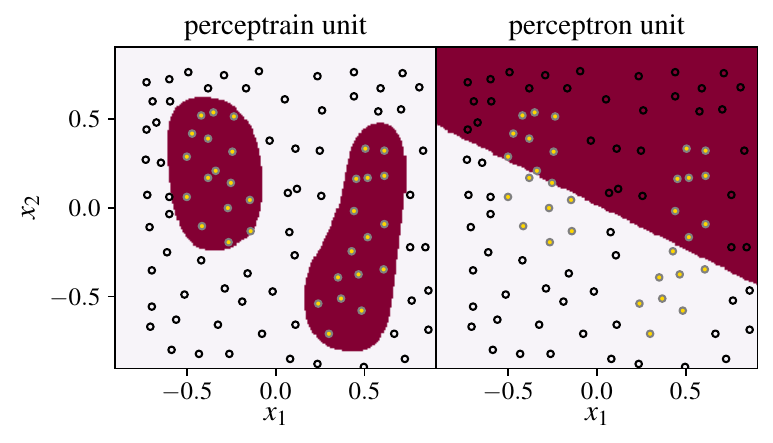}
    \caption{A standard perceptron defines a hyperplane in the space of inputs (right), while the perceptrain (left) defines a more complex boundary. Empty (full) symbols represent the data for the first (second) class of object to be classified while the shaded region corresponds to the result of the classification after training.
    \label{fig:Decision-boundary}}
\end{figure}

As a first (toy) illustration of the perceptrain, we have used a single perceptron and a single perceptrain to perform a simple classification task in two dimensions (with artificial data). The results are shown in Fig.~\ref{fig:Decision-boundary} for a small bond dimension $\chi=5$ and $d=5$ with the symbols representing the data (filled or empty depending on the class) and the shaded regions the classification resulting from the perceptrain (left) and the perceptron (right). The comparison is, of course, not fair to the perceptron since they are seldom used outside of a network and the perceptrain has many more parameters. The number of parameters of the perceptrain scales as $\chi^2 n d$ which reduces to $\chi d$ in our $n=2$ two dimensional toy example. This should be compared to e.g. a simple single-layer perceptron network whose number of parameters scales as $ln$, with $l$ being the depth of the layer. The point of this figure is to illustrate that a single perceptrain already has a large expressing power, comparable to the one of a small network. 

\subsection{Discussion}

From a purely mathematical perspective, neural networks support two crucial features: \emph{evaluation} (for a given $\boldsymbol{x}$, one can easily compute $\phi(\boldsymbol{x})$) and \emph{automatic differentiation} (through the celebrated backward propagation algorithm, the cost to obtain the full gradient $\partial\phi(\boldsymbol{x})/\partial\boldsymbol{W}$ is roughly the same as one evaluation). At a more qualitative level, the fact that the perceptron is bio-inspired may make it well adapted to represent the type of data that brains can process.

The perceptrain, however, has the following set of properties that we will leverage on in the rest of the article:
\begin{itemize}
\item Evaluation is straightforward and amounts to $n$ matrix-vector multiplications for a total complexity $O(n \chi^2)$.
\item Automatic differentiation is also straightforward. The gradient with respect to all matrices $A_i^s$ can be computed for the same cost as the evaluation provided that (as in the backward propagation algorithm) one caches the intermediate matrix vectors product during the evaluation:
\begin{eqnarray}
\frac{\partial\phi(\boldsymbol{x})}{\partial [A_i^s]_{pq}} &=
T_s(x_i) 
\beta [1-\phi(\boldsymbol{x})^2]
[A_1(x_1)...A_{i-1}(x_{i-1})]_{p} \nonumber \\   
&[A_{i+1}(x_{i+1})...A_n(x_n)]_{q}
\end{eqnarray}
\item Local updates. One of the properties at the origin of the success of MPS is that, even though a matrix $A_i^s$ is "local" in the sense of being associated with one variable $s_i$, modifying it has a global impact on the full function. Algorithms such as DMRG only make local updates but nevertheless are capable of finding global minima very efficiently \cite{schollwoeck_density-matrix_2005}.
\item Compression. An MPS can be efficiently compressed to lower the rank $\chi$ while losing a minimum of information. This is done by performing a series of singular value decomposition (SVD) for each matrix $A_i^s$ and truncating the matrices by ignoring the smallest singular values.
\item Dynamic rank increase. Conversely, it is possible to dynamically increase the rank of an MPS so that it becomes gradually more expressive. This can be done by fusing two consecutive matrices introducing $B_i^{ss'} = A_i^s A_{i+1}^{s'}$ and performing the optimization on $B_i^{ss'}$ instead of $A_i^s$ and $A_{i+1}^{s'}$. In a second stage the MPS structure is recovered by performing an SVD of $B_i^{ss'}$, increasing the rank in the process.
\item Initialization. The recent development of the Tensor Cross Interpolation algorithm \cite{dolgov_parallel_2020,nunez_fernandez_learning_2022,nunez_fernandez_learning_2025} allows one to initialize an MPS with a large class of functions. 
\item Quantum inspired instead of bio inspired. Last, MPS are natural structures to express many-body wave functions, so the perceptrain may be well adapted for VMC.
\end{itemize}

In the rest of this article, we use the perceptrain to build simple yet efficient many-body ansatz for VMC. We leave the exploration of its usefulness for other tasks to future works.

\section{Simple many-body ansatz with perceptrains}
\label{subsec:Perceptrain-for-spin-models}

In this section, we construct a simple two-layer neural network of perceptrains, referred to as "perceptrain network" (\perceptrainn), that we shall use as a variational ansatz. It is a direct extension of MPS to address, in particular, two-dimensional systems. For the sake of concreteness, we focus on quantum spin models but the content of this section applies to fermionic models as well.

\begin{figure}[t]
    \centering
        \includegraphics[width = 1.0\linewidth]{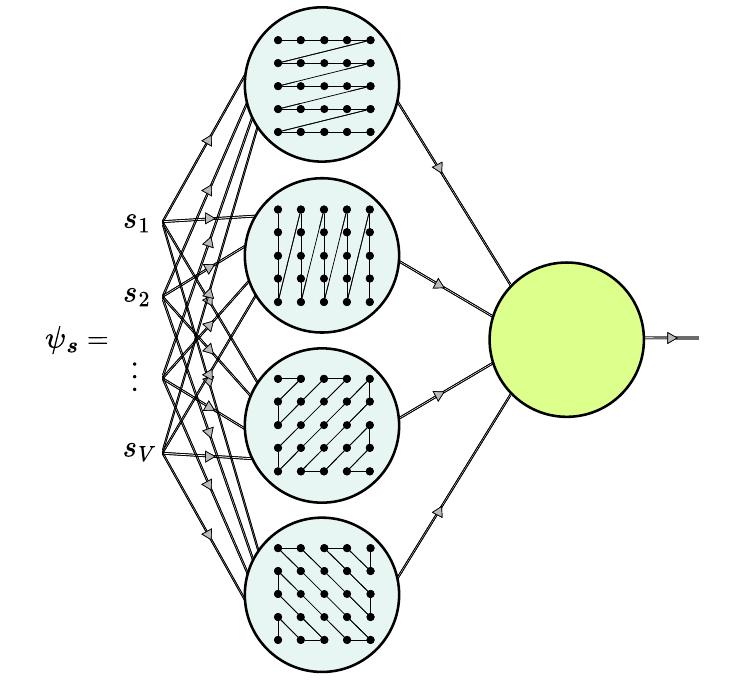}
    \caption{A representation of the \perceptrain ansatz Eq.~\eqref{eq:vwf-perceptrain} based on the schematics of Fig.~\ref{fig:drawing-perceptrain}. The four child perceptrains $\phi_{c}^{a}$ (blue) all get the same configuration $\boldsymbol{s}$, and feed their output to the parent perceptrain $\phi_{p}$ (green) which outputs the value of the wave function. The four orderings of the tensors in $\phi_{c}^{a}$ (formally, the four different permutations $\boldsymbol{\sigma}$) are shown inside the blue circle for a $5\times 5$ spin lattice. These horizontal, vertical, right diagonal and left diagonal orderings are the actual orderings used in this work.
        \label{fig:Paths-drawing}}
\end{figure}

\subsection{Variational Monte Carlo framework}
\label{subsec:VMC}

We consider a configuration space that consists of $V$ 1/2-spins $\boldsymbol{s}= (s_1,...s_V)$ with $s_i\in \pm 1$. We seek the ground state of a Hamiltonian $H$ that can be very general; the only requirement at this stage is that for a given $\boldsymbol{s}$, there exists only a small (typically $O(V)$) number of configurations $\boldsymbol{s'}$ for which $H_{\boldsymbol{s}\boldsymbol{s'}}\ne 0$. We construct a (unnormalized) variational wave function $\psi_{\boldsymbol{s}}=\psi_{\boldsymbol{s}}(\boldsymbol{\theta})$ that depends on a vector of parameters $\boldsymbol{\theta}$ and seek to minimize its energy
\begin{equation}
E(\boldsymbol{\theta}) = 
\frac{ \sum_{\boldsymbol{s}\boldsymbol{s'}} \psi_{\boldsymbol{s}}^* 
H_{\boldsymbol{s}\boldsymbol{s'}} \psi_{\boldsymbol{s'}}}{\sum_{\boldsymbol{s''}} |\psi_{\boldsymbol{s''}}|^2}.
\end{equation}
Introducing the \emph{local energy}
\begin{equation}
e_{\boldsymbol{s}} = 
\frac{ 1}{\psi_{\boldsymbol{s}}}
\sum_{\boldsymbol{s'}} H_{\boldsymbol{s}\boldsymbol{s'}} \psi_{\boldsymbol{s'}},
\end{equation}
the energy can be recast into
\begin{equation}
E(\boldsymbol{\theta}) = 
\sum_{\boldsymbol{s}}
\frac{  |\psi_{\boldsymbol{s}}|^2}{\sum_{\boldsymbol{s''}} |\psi_{\boldsymbol{s''}}|^2}
e_{\boldsymbol{s}} 
= \langle e_{\boldsymbol{s}} \rangle_{|\psi|^2},
\end{equation}
where the second equality means that the local energy needs to be averaged with respect to the probability distribution $p_{\boldsymbol{s}} = |\psi_{\boldsymbol{s}}|^2 / \sum_{\boldsymbol{s''}} |\psi_{\boldsymbol{s''}}|^2$. The appeal of this formulation comes from the fact that $p_{\boldsymbol{s}}$ can be sampled through Markov chain Monte Carlo (e.g. Metropolis algorithm) and the corresponding energy estimated as $E(\boldsymbol{\theta}) \approx (1/M) \sum_{\alpha=1}^M e_{\boldsymbol{s_\alpha}}$ where $M$ is the number of samples \cite{becca_quantum_2017}. A typical value used in most of this work is $M=3200$. A very nice property of VMC is that as $\psi$ approaches the actual ground state $\psi^0$ of $H$ (with energy $E_0$), the local energy approaches $e_{\boldsymbol{s}} \approx E_0$, i.e. the local energy does not depend on the configuration anymore. It follows that the error of the estimate of the energy, at fixed $M$, decreases upon improving the wave function. Ideally, if one could construct $\psi^0$ exactly, a single configuration $\boldsymbol{s}$ would be sufficient to obtain the energy exactly (in contrast to the quantum computer counterpart of VMC \cite{louvet_feasibility_2024}).
 
\subsection{Ansatz construction}
\label{subsec:PN-ansatz}

We now proceed with the construction of the \perceptrain ansatz. The construction was motivated by remarking that MPS capture local correlations in 1D problems very well, because the MPS's structure reflects the actual geometry of the problem. In 2D, the natural generalization is PEPS but the corresponding computational overhead quickly becomes prohibitive. We choose instead to construct a network that involves several MPS, each associated to a different ordering of the spins, so that spins that are close spatially are also close in the ansatz for at least one of those MPS (see Fig.~\ref{fig:Paths-drawing}).

More precisely, we use $K$ different MPS to build $K$ different "child" perceptrains $\phi^a_c$, for $a\in\{1...K\}$ (typically $K=4$ in our numerics). Each child perceptrain is described by the $n$ matrices $A_i^s$ and a permutation $\boldsymbol{\sigma}$ of the spins (the "ordering" of the spins) such that $\boldsymbol{\sigma}_i = j$ with $i,j\in\{1...V\}$. We implicitly assume the index $a$ in these objects, to avoid cumbersome notation. $\phi^a_c ({\boldsymbol{s}})$ then takes the form 
\begin{equation}
\phi^a_c ({\boldsymbol{s}}) = \mathbf{tanh}[b_{c}+\beta_{c}  \mathbf{Tr} 
A_1^{s_{\boldsymbol{\sigma}_1}} A_2^{s_{\boldsymbol{\sigma}_2}}...A_n^{s_{\boldsymbol{\sigma}_V}}].
\end{equation}
The full network consists of $K$ "child" perceptrains of rank $\chi$ and one "parent" (continuous) perceptrain $\phi_p$ of rank $\chi^p$. The variational wave function reads
\begin{equation}\label{eq:vwf-perceptrain}
\psi_{\boldsymbol{s}} = \phi_p ( \phi^1_c({\boldsymbol{s}}),\phi^2_c({\boldsymbol{s}})...\phi^K_c({\boldsymbol{s}}) ).
\end{equation}
We emphasize that evaluating $\psi_{\boldsymbol{s}}$ amounts to a few MPS contractions. It is therefore computationally cheap with a cost that scales as $K V \chi^2$.  Here, the two-dimensional nature of the problem has been incorporated using the different orderings of the sites (as shown Fig.~\ref{fig:Paths-drawing}), circumventing paying the exponentially large cost coming from contracting a genuine two-dimensional tensor network.

In the calculations performed in this manuscript, we worked with only two values of $\chi^p$, $\chi^p=2$ (\perceptrainn) and $\chi^p=1$ (SBS limit). We call the case $\chi^p=1$ the "SBS limit" since the corresponding \perceptrain ansatz reduces to the SBS ansatz developed previously \cite{schuch_strings_2008,sfondrini_simulating_2010,glasser_neural-network_2018}, provided we choose $b_c=0$, and $\beta_{c}$ to be small enough for the non-linearities to be ignored. We found empirically, for our studied system, that the increase of accuracy obtained when going from $\chi^p=1$ to $\chi^p=2$ was relatively modest and it was more effective to increase $\chi$, or, even better, $K$. When setting $K=1$ (the MPS limit), we chose the horizontal ordering (top one in Fig.~\ref{fig:Paths-drawing}). In the SBS limit, the main difference between our approach and the one of \cite{schuch_strings_2008,sfondrini_simulating_2010} is that we perform the optimization \emph{locally} as we shall discuss next.

\subsection{Optimization procedure}
\label{subsec:VMC-optimisation}

In the spirit of the two-site DMRG algorithm, we perform the optimization by iterating over the different matrices of the different MPS, optimizing two consecutive matrices at a time. For each pair of matrices $A_i$ and $A_{i+1}$ the algorithm is the following:

\begin{enumerate}[label={(\arabic*)}]
\item put the orthogonal center of the corresponding MPS at the position $i$. See Refs.~\cite{schollwoeck_density-matrix_2005,schollwock_density-matrix_2011} for a discussion of orthogonal centers in MPS. Fixing the orthogonal center provides several advantages: first it "brings" information from other matrices down to the ones that are being optimized, thereby conferring the local optimization a more global nature; second, it provides a natural metric for the tensors being optimized since their $L_2$ norm becomes the one of the full tensor (in that sense the corresponding gradient would be a natural gradient for a single MPS). Third, it fixes the "gauge" freedom inherent to an MPS.
\item fuse $A_i^s$ and $A_{i+1}^{s'}$ into $B_i^{ss'} = A_i^s A_{i+1}^{s'}$.
\item Generate $M$ new samples by performing a few, typically one or two, Metropolis moves for each sample. These samples may be correlated with the previous ones but are not correlated between themselves, which is an important feature.
\item calculate the gradient $\partial E/\partial B_i$. This gradient can be estimated from the samples through,
\begin{equation}\label{eq:gradient-energy-stochastic}
\partial_{\boldsymbol{\theta}}E = 2 [\langle e_{\boldsymbol{s}}\mathcal{O}^{\boldsymbol{\theta}}_{\boldsymbol{s}}\rangle_{|\psi|^2}-E\langle\mathcal{O}^{\boldsymbol{\theta}}_{\boldsymbol{s}}\rangle_{|\psi|^2}],
\end{equation}
where $\mathcal{O}^{\boldsymbol{\theta}}_{\boldsymbol{s}}=\partial_{\boldsymbol{\theta}}\psi_{\boldsymbol{s}}/\psi_{\boldsymbol{s}}$ is the logarithmic derivative of $\psi$ at site $\boldsymbol{s}$, over parameters $\boldsymbol{\theta}=B_i$. 
\item optimize $B_i$ using some sort of gradient descent. Here we use two different options: either a normalized gradient descent,
\begin{equation}\label{eq:gradient-descent}
\boldsymbol{\theta} \to \boldsymbol{\theta} - \alpha \frac{\partial_{\boldsymbol{\theta}}E}{\|\partial_{\boldsymbol{\theta}}E\|_2}
\end{equation}
where $\alpha$ is the learning rate. Alternatively, we can use the so-called stochastic reconfiguration (SR) \cite{sorella_green_1998,sorella_green_2000,casula_geminal_2003} update which corresponds to an imaginary time evolution projected within the variational subspace or equivalently to natural gradient descent with respect to the quantum geometric tensor \cite{filippi_optimal_2000},
\begin{equation}\label{eq:Geometric-tensor}
S_{ij}=\langle\mathcal{O}_{\boldsymbol{s}}^{\theta_i}\mathcal{O}_{\boldsymbol{s}}^{\theta_j}\rangle_{|\psi|^2}-\langle\mathcal{O}_{\boldsymbol{s}}^{\theta_i}\rangle_{|\psi|^2}\langle\mathcal{O}_{\boldsymbol{s}}^{\theta_j}\rangle_{|\psi|^2}.
\end{equation}
The SR update is
\begin{equation}\label{eq:SRequation}
\boldsymbol{\theta} \to \boldsymbol{\theta} - \alpha S^{-1} \partial_{\boldsymbol{\theta}}E.
\end{equation}
In practice, the geometric tensor can become singular and must be regularized \cite{roth_high-accuracy_2023}. This is done by introducing a shift of its diagonal $S_{ii} \to S_{ii} + \epsilon_1S_{ii}+\epsilon_2$. Typical values of the regularization parameters that we used are $\epsilon_1=0.01$ and $\epsilon_2=0.001$. In practice, the $S$ matrix is not inverted, rather one solves for the corresponding linear problem. This step may be computationally costly when one performs the optimization with respect to all the parameters of the ansatz (see \cite{chen_empowering_2024} for a mitigation strategy). However, in the present case, since we only optimize with respect to a small subset of parameters, this bottleneck is absent.
\item perform an SVD of $[B_i^{ss'}]_{aa'}$ considered as a matrix with $a,s$ indexing its rows and $a',s'$ indexing its columns, 
\begin{equation}
[B_i^{ss'}]_{aa'} = \sum_b U_{ab}^{s} \lambda_b V_{ba'}^{s'},
\end{equation}
truncate $\lambda_b$ to keep only the $\chi$ largest singular values and update $[A_i^s]_{ab}\to U_{ab}^{s}$ and $[A_{i+1}^{s'}]_{ba'} \to \lambda_b V_{ba'}^{s'}$. This step can be used to increase (or possibly decrease) the rank of the MPS.
\end{enumerate}
The above sequence is repeated for all MPS and all adjacent pairs of matrices within an MPS. This forms a ``sweep.'' The precise order in which these pairs are optimized does not appear to be very relevant in our experience. In our numerical experiments for the "2D model" (see Sec.~\ref{sec:Application-to-Spin-models}), convergence of the optimization is usually reached within $\sim 150$ sweeps while $333$ sweeps form what we call one ``Epoch.''

We have performed three different variants of the above optimization which we tag "static," "dynamical-$\chi$," and "dynamical-$K$" optimization, the merit of which will be discussed below. They are defined as
\begin{itemize}
\item "static" means that we fix $K$ ($K=4$) and $\chi$ at the beginning of the calculation and optimize all parameters one after another.
\item "dynamical-$\chi$" means that we start with a low value of $\chi$, optimize the parameters, increase $\chi$ [following step (6)] and then repeat the sequence until the targeted value of $\chi$ has been reached. 
\item "dynamical-$K$" means that we start with all the children perceptrains initialized with the constant function and optimize only the $K=1$ child perceptrain, then start to optimize the second child perceptrain (considering the first child as fixed), and continue the sequence until all child perceptrains have been optimized. 
\end{itemize}

\subsection{Scaling of the algorithm}

We summarize here the scaling of the different parts of the algorithm. Evaluating an MPS for a single configuration takes $n$ matrix-vector multiplications, and hence scales as $\mathcal{O}(\chi^2 n)$. The memory footprint is  $2\chi^2 n$. Putting an MPS in the canonical form uses the QR decomposition and therefore scales as $\mathcal{O}(\chi^3 n)$. Taking the derivative of an MPS with respect to one tensor scales as one evaluation (MPS naturally implements automatic differentiation). Last, during an evaluation, one can cache the left and right partial contractions (the "environments" of a site) making it possible to compute a new configuration where one spin is flipped with only $\mathcal{O}(\chi^2)$ (for the cost of doubling the memory footprint).

In the present context, the above translates into a single evaluation of $\psi(\boldsymbol{s})$ that scales as $\mathcal{O}(KV\chi_{c}^{2})$. The cost of evaluating the ``parent'' perceptrain in our examples is negligible, and can be ignored. The rest of the scaling of the approach is Hamiltonian dependent and we specialize here to the transverse field Ising model described in the next section. The calculation of the local energy requires, for each configuration $\boldsymbol{s}$, to calculate $\psi(\boldsymbol{s'})$ for all $\boldsymbol{s'}$ such that $H_{\boldsymbol{s's}}\ne 0$. In the present case, there are $V$ such neighbors. This in principle requires paying an additional factor $V$. However, using the caching trick mentioned above, this factor can be avoided. Since we consider a long range model, the calculation of the diagonal energy itself  $H_{\boldsymbol{ss}}$ scales as $V^2$. Overall, for $N_\text{opt}$ optimization steps and $M$ samples, we arrive at a time complexity that scales as
\begin{equation}
CPU \sim N_\text{opt}  \left( M KV\chi_{c}^{2} + M V^2 + KV\chi_{c}^{3}\right)
\end{equation}
where in practice, the computing time is entirely dominated by the first term in the above equation. The memory footprint, however, consists of the $M$ configurations ($MV$ bits),  the ansatz itself and a (optional) last term that corresponds to the caching, and dominates in practice
\begin{equation}
MEM \sim MV + KV\chi_{c}^{2} + MKV\chi_{c}^{2}.
\end{equation}
For the typical numbers used in this work, the memory footprint is very mild, even using the caching.

\section{Application to Transverse field spin model}
\label{sec:Application-to-Spin-models}

\begin{figure*}[t]
    \centering
        \includegraphics[width = 1.0\linewidth]{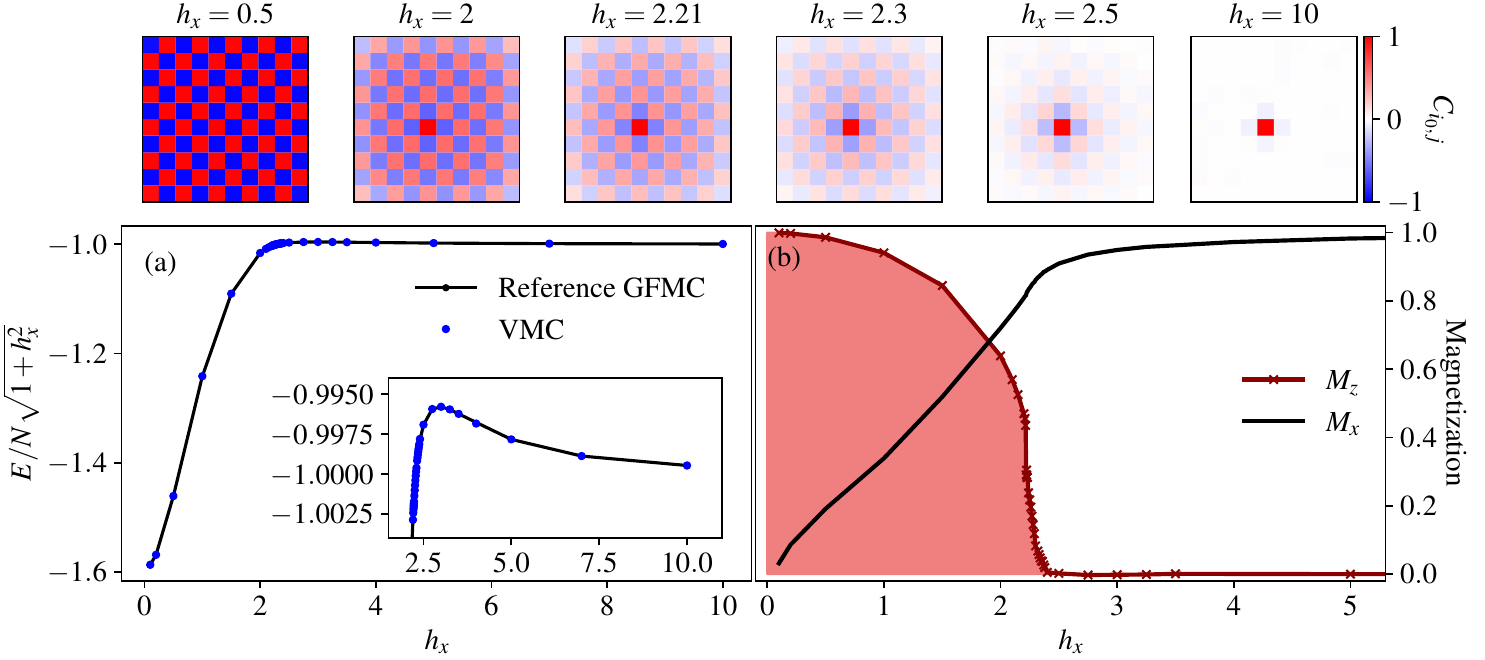}
    \caption{Phase diagram of the 2D model on a $10\times 10$ lattice. (Upper panel) The correlation function $C_{i_0j}$ between the site $i_0 = (i_x=5, i_y=5)$ and all the other sites $j$ on the lattice for different values of $h_{x}$. (a) Energy as a function of $h_{x}$. The ground state energy obtained by VMC (blue points) is compared to the reference energies (black line) calculated using GFMC algorithm (see Sec.~\ref{subsec:GFMC}). (b) Staggered magnetization along the $z$ direction (red) and magnetization along the $x$ direction (black) as a function of $h_{x}$.
        \label{fig:Physics}}
\end{figure*}

We now turn to our numerical calculations where we focus on a transverse field Ising model in two dimensions. This model is interesting for its relative tractability, its link with quantum annealing experiments with Rydberg atoms and the absence of sign problem which allows us to verify the quality of our results using the post-VMC Green function Monte Carlo method (GFMC).

A rule of thumb with variational techniques is that a relatively accurate energy can correspond to very inaccurate physics. This is due to the energy being mostly controlled by relatively local physics while important long range behavior are associated with relatively small energy variations. It follows that a relatively poor variational ansatz may provide a correct equation of states but that one should be very careful with other observables than the energy. We shall illustrate this principle in our calculations with typically relative precisions better than $10^{-3}$ being needed to correctly capture the phase diagram. We shall also find that the accuracy correlates very well with the so-called "$V$-score," which has been introduced recently \cite{wu_vscore_2024}.

\subsection{Model and its phase diagram.}
\label{subsec:model}

We consider the transverse field quantum Ising model with $V$ sites. The Hamiltonian reads
\begin{equation}\label{eq:IsingRydberg}
H=\sum_{i<j}J_{ij}\sigma^z_i\sigma^z_j-h_x\sum_i\sigma_i^x.
\end{equation}
where $\sigma^z$ and $\sigma^x$ are the Pauli matrices and $J_{ij}$ the exchange interaction matrix. A configuration $\boldsymbol{s}=(s_1,s_2,\dots,s_V)$ corresponds to the values of the spins along the $z$ axis so that
\begin{equation}
H_{\boldsymbol{s}\boldsymbol{s}} = \sum_{i<j}J_{ij}s_is_j
\end{equation}
and for $\boldsymbol{s}\ne \boldsymbol{s'}$,
\begin{equation}
H_{\boldsymbol{s}\boldsymbol{s'}} = -h_x \sum_{a=1}^V
...
\delta_{s_{a-1}s_{a-1}'}
\delta_{s_{a},-s_{a}'}
\delta_{s_{a+1}s_{a+1}'}
...
\end{equation}
We perform almost all our calculations on a two dimensional system of $V=10\times 10$ spins with open boundary conditions and an untruncated antiferromagnetic $J_{ij} = 1/r^6$ interaction where $r$ is the Euclidean distance between $i$ and $j$ on a square lattice with unit lattice spacing. In this setup the second nearest neighbor interacting term is $1/\sqrt{2}^6 = 1/8$ of the nearest neighbor term, making this model mildly frustrated. The model can be directly mapped to the Rydberg atom quantum annealing model \cite{samajdar_complex_2020,scholl_quantum_2021,ebadi_quantum_2021,kalinowski_bulk_2022} using a simple change of variables. We refer to this model as the "2D model." For benchmark purposes, we also considered a simpler exactly solvable one dimensional version, the "1D model" of $V=10$ sites with the interaction limited to nearest neighbors and periodic boundary conditions. Both models exhibit a quantum phase transition at finite transverse field $h_x$ between an ordered phase and a paramagnetic phase \cite{kalz_phase_2008}. Besides the energy, we will compute the following physical observables: the correlation function, the total staggered magnetization along the $z$ direction, and the total magnetization along the $x$ direction,
\begin{eqnarray}
\label{eq:Correlation}
C_{ij} &=&\langle \sigma^z_i\sigma^z_j \rangle = \langle s_i s_j \rangle_{|\psi|^2},
\\
\label{eq:Stagg_mag}
M_{z} &=& \frac{1}{V}\sum_{i =(i_x,i_y)} (-1)^{i_x+i_y} \langle \sigma_{i}^{z}\rangle,
\\
\label{eq:x_mag}
M_{x}&=& \frac{1}{V}\sum_{i =(i_x,i_y)} \langle \sigma_{i}^{x}\rangle.
\end{eqnarray}

\begin{table*}[t]
\caption{\label{tab:VMC-extrapolation}
Ground state energy per site for the 2D model.
 }
\begin{ruledtabular}
\begin{tabular}{l D{.}{.}{5} D{.}{.}{5} D{.}{.}{5} D{.}{.}{5} D{.}{.}{5} D{.}{.}{5}}
$h_{x}$
& 		10 		 &		 3		& 	\text{2.5}	        & 	\text{2.3}	       & 	\text{1.5} 	      & 	    1\\
\colrule
VMC
& \textbf{-10.0444}17 & \textbf{-3.1489}41 & \textbf{-2.6842}42 & \textbf{-2.50}6990 & \textbf{-1.9662}35 & \textbf{-1.756}279 \\
\colrule
\multicolumn{1}{c}{Extrapolation\footnote{Linear var$[H] / E^{2}\to0$ extrapolation.}} 
& \textbf{-10.04442}6 &  \textbf{-3.1489}71& \textbf{-2.6842}78 & \textbf{-2.5070}70 & \textbf{-1.9662}71 &  \textbf{-1.756}306 \\
$\;\;\;\pm$
& \text{0.0}00001       & \text{0.0}00025      &  \text{0.0}00007    & \text{0.0}00005     &     \text{0.0}00009 &  \text{0.0}00010 \\
\colrule
GFMC
& \textbf{-10.04442}6& \textbf{-3.1489}59  & \textbf{-2.68426}6 & \textbf{-2.5070}60 & \textbf{-1.9662}60 & \textbf{-1.756}298\\
$\;\;\;\pm$
& \text{0.0}00001      & \text{0.0}00003      & \text{0.0}00003      & \text{0.0}00007     & \text{0.0}00002     & \text{0.0}00002\\
\colrule
MPS\footnote{With $\chi=16$}
& \textbf{-10.04}371 & \textbf{-3.1}33674  & \textbf{-2.6}64684   &  			       & \textbf{-1.96}5813 & \textbf{-1.75}4742\\
\end{tabular}
\end{ruledtabular}
\end{table*}

\begin{figure}[b]
    \centering
        \includegraphics[width = 1.0\linewidth]{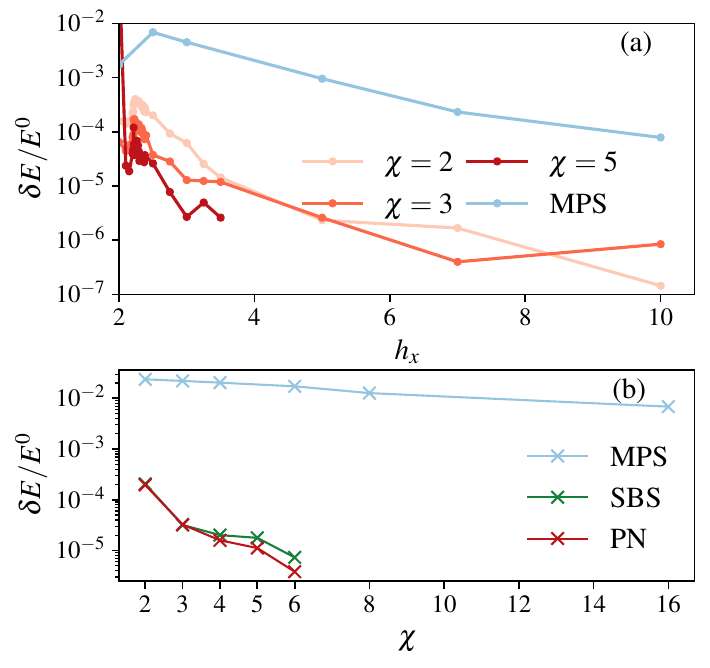}
    \caption{Comparison between the perceptrain-based ansatz and MPS of $\chi=16$ in the 2D model. (a) Relative error in the energy as a function of $h_{x}$ for different values of $\chi$. (b) Relative error in the energy as a function of $\chi$ at $h_x=2.5$, close to the critical point.
    \label{fig:2D-Expressivity}}
\end{figure}

Before exploring different aspects of our ansatz and optimization procedure, we quickly survey the final results that we obtained for the 2D model. Essentially, we find that, using our perceptrain based ansatz, we could \emph{solve} the 2D model with an accuracy good enough to actually obtain the entire phase diagram, including the quantum critical point, in a reliable way. The results are shown in Fig.~\ref{fig:Physics}.

Figure~\ref{fig:Physics}(a) shows the obtained VMC energy versus $h_x$ compared with our reference GFMC energy. At large field, $E(h_x)\to -N h_x$ (all spins polarized along the $x$ axis) and at zero field $E(0)\approx -1.5933 N$ (classical limit). For the energy to remain within a narrow range as we vary $h_x$ (so that features may be more visible), we plot the energy in the units of $N \sqrt{1+h_x^2}$. We find a perfect agreement between VMC and GFMC in the whole range $h_x\in [0,10.]$ (see Table~\ref{tab:VMC-extrapolation} for a more quantitative statement). Zooming around $E/(N \sqrt{1+h_x^2})\approx -1$, we see that the agreement remains up to at least the fourth digits.

Figure~\ref{fig:Physics}(b) shows the staggered magnetization $M_z$ and the transverse magnetization $M_x$ versus $h_x$. $M_z$ acquires a finite value for $h_x< 2.3$ where the system becomes antiferromagnetic. Conversely, $M_x$ saturates in the paramagnetic phase for $h_x> 2.3$ and is linear in the antiferromagnetic phase. The behavior of $M_x$, however, is much less sharp than the one of $M_z$. Here we computed $M_x$ directly but it is fully consistent with the data of Fig.~\ref{fig:Physics}(a) using $M_x = \partial_{h_{x}}E$ and e.g. finite difference derivation.

The upper panels of Fig.~\ref{fig:Physics} show the correlation function $C_{ij}$. We observe, as expected, the development of strong antiferromagnetic correlations around the critical point in the window $h_{x}\in [2.2, 2.5]$. The goal of this article is the analysis of the capabilities of our perceptrain based ansatz, hence we shall stop here the study of the physics of the 2D model. It seems that our approach may be used as a very precise "simulated quantum annealer," but we differ the corresponding study to a subsequent publication. 

\subsection{Perceptrain-based ansatz versus MPS in 2D}
\label{subsec:2D-perceptrain-vs-MPS}

One of the chief motivations for our ansatz was to bypass the limitations of MPS ansatz in two dimensions. It is therefore crucial to show that, indeed, the perceptrain-based ansatz performs better than a simple MPS for the 2D model. The results are shown in Fig.~\ref{fig:2D-Expressivity}(a). We find that we can consistently obtain relative accuracies better than $10^{-4}$ for very low bond dimensions $\chi \in [3-5]$ while regular MPS struggle to reach the $10^{-2}$ accuracy level even using larger bond dimensions. This is a very strong point in favor of our ansatz. Note that in DMRG, it is very common to use values of $\chi \sim 1000$, or even larger. Obtaining such high level of precision with such low bond dimension is a clear indication that the ansatz is particularly adapted to the 2D model. As expected, at a fixed rank (i.e. fixed level of expressivity), the precision decreases close to the criticality. However, increasing the rank consistently leads to higher and higher precision, allowing us to improve, if higher precision be desired. Figure~\ref{fig:2D-Expressivity}(b) shows that our error level quickly improves with $\chi$ as opposed to the corresponding MPS behavior. We also find that, in this regime, the SBS limit is almost as good as the full-fledged perceptrain based ansatz. Within our current setup, our accuracy appears to be limited to the $\sim 10^{-7}$ level which we attribute to the remaining noise present using $M=3200$ walkers.

\begin{figure}[t]
    \centering
        \includegraphics[width = 1.0\linewidth]{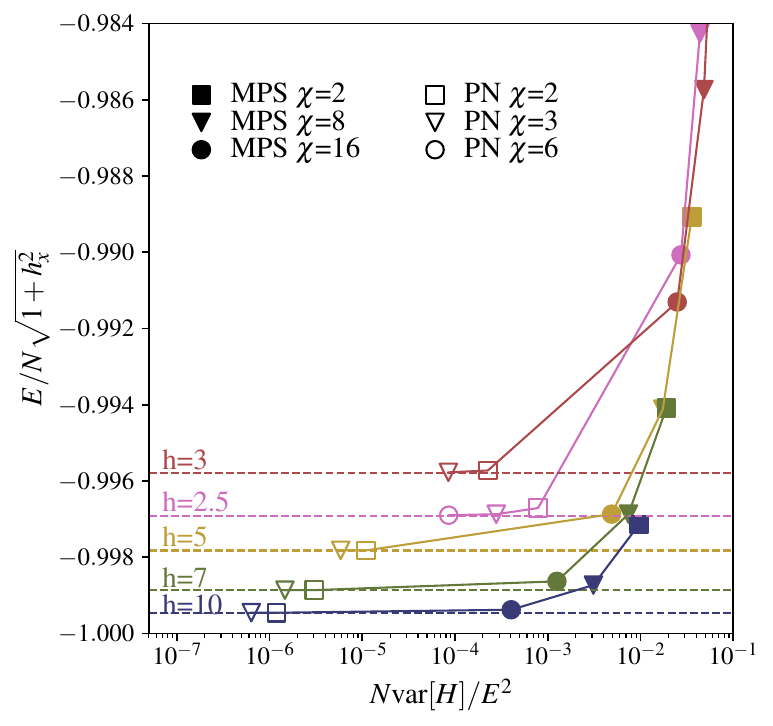}
    \caption{
     Best obtained energy as a function of the $V$-score $N$var$[H]/E^{2}$ for different values of $h_{x}$ in the 2D model (different colors). MPS (filled symbols) is not able to describe nontrivial ground states at lower values of $h_{x}$ even for $\chi=16$. At the same time, the \perceptrain (empty symbols) leads to good precision already at $\chi=2$. At $h_{x}=2.5$, close to the critical point, we obtain another digit in energy precision simply by increasing the rank from $\chi=2$ to $\chi=6$.
    \label{fig:2D-Comparison-models}}
\end{figure}

Figure~\ref{fig:2D-Comparison-models} shows the same data but we plot the energy $E$ as a function of the variance of the energy   var$[H] = \langle H^2\rangle - E^2$. The variance is rescaled with respect to the size of the system and an estimate of the bandwidth of the energy spectrum, the so-called $V$-score \cite{wu_vscore_2024} which for our model evaluates to $V$-score $= N \mathrm{var} H/E^2$. As we shall see the $V$-score is a very good proxy for the accuracy of the calculation. $V$-scores in the $10^{-6}-10^{-4}$, as obtained here, are considered of very high quality. This figure shows once again the large difference of ansatz quality between the MPS and the perceptrain ansatz (except for large value of $h_x$ where the ground state eventually converges to a product state, i.e. a rank one MPS). We could, of course, push the rank of the MPS to much higher values. The point here is that with the perceptrain ansatz, an extremely low rank $\chi=2$ or $3$ is sufficient to describe the ground state.

\begin{figure}[t]
    \centering
            \includegraphics[width = 1.0\linewidth]{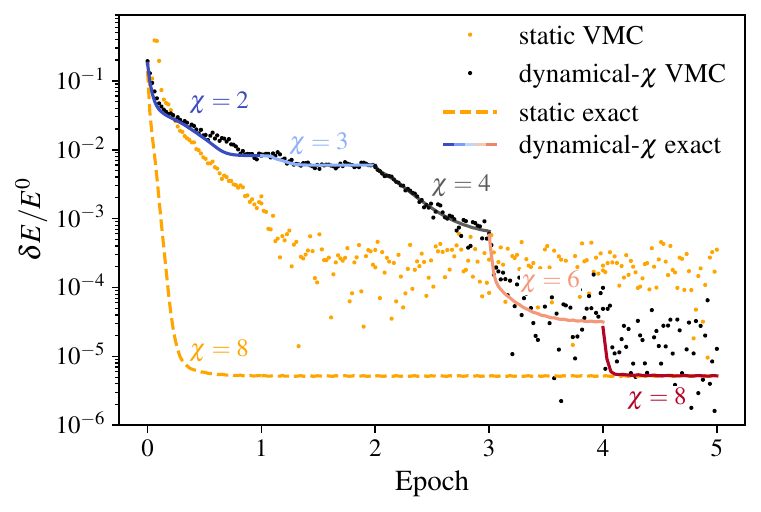}
    \caption{
    Relative error in energy along the optimization of the 1D model [Eq.~\eqref{eq:IsingRydberg}] at $h_{x}=1$ for two optimization methods: static (various colors) and dynamical-$\chi$ (gold). The gradient of the energy was either calculated exactly (lines) or in a nosy way using VMC (dot symbols, the only available method for large systems).  The stochastic dynamical-$\chi$ optimization is robust to the presence of VMC noise while the static optimization is not (much slower descent and it gets stuck at a much lower precision). 
        \label{fig:Exact-VMC}}
\end{figure}

\begin{figure}[b]
    \centering
        \includegraphics[width = 1\linewidth]{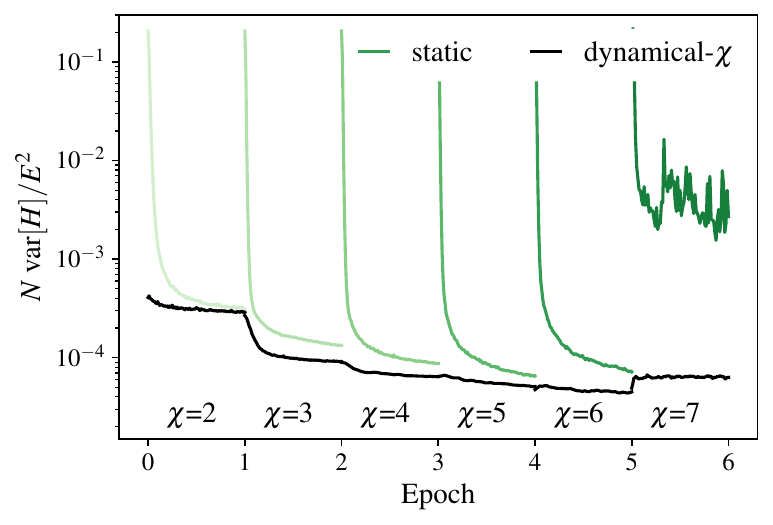}
    \caption{
   $V$-score along the dynamical-$\chi$ VMC optimization (black) compared to the static VMC optimization (green) at corresponding $\chi$, for $h_{x}=3$. The dynamical-$\chi$ optimization is more stable than static. $\alpha=10^{-3}$.
    \label{fig:Breaking-of-optimisation1}}
\end{figure}

\subsection{Effect of the optimization path on the final precision}
\label{subsec:1D-optimisation-path}

One of the hypotheses of this article is that the optimization path crucially affects the final accuracy that will be reached. More precisely, we surmise that the ability to dynamically increase the rank of the child perceptrains during optimization will make the optimization process more robust.

\begin{figure*}[t]
    \centering
        \includegraphics[width = 1.0\linewidth]{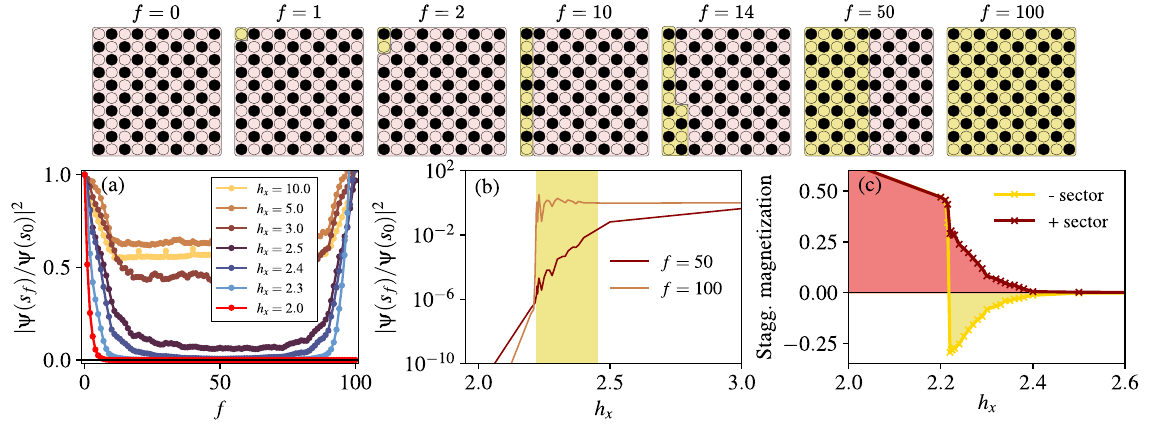}
    \caption{Evolution of the variational wave function $\psi$ along the domain wall path $\boldsymbol{s}_{f}$. (Upper panel) A few explicit configurations $\boldsymbol{s}_{f}$. Filled circles correspond to up spins, empty circles to down spins; the color (red and yellow) identifies the two equivalent ground states with a domain wall in between. (Lower panels) (a) Probability to be in $\boldsymbol{s}_{f}$, normalized by the probability to be in $\boldsymbol{s}_{0}$ for different values of $h_{x}$. At small values of $h_{x}$, the $\boldsymbol{s}_{V/2}$ becomes so improbable, that walkers stay trapped in their respective sectors. (b) Probability of $\boldsymbol{s}_{V/2}$ and $\boldsymbol{s}_{V}$ as a function of $h_{x}$.  (c) Hysteresis in $M_{z}$ versus $h_x$. For each value of $h_x$, a single VMC optimization has been performed; the corresponding wave function is sampled in two different ways by initializing all the walkers in either the positive ($\boldsymbol{s}_{0}$, red) or the negative ($\boldsymbol{s}_{V}$, yellow) sector.
         \label{fig:Wave-function-barrier}}
\end{figure*}

We now examine this hypothesis on our data. First, Fig.~\ref{fig:Exact-VMC} shows two optimization procedures for the 1D model: a static optimization where the rank is fixed at $\chi=8$ at the beginning and a dynamical-$\chi$ where we start at $\chi=2$, and increase the rank, until $\chi=8$ is reached. Since the system is small ($V=10$ sites) we can also calculate the exact gradient of the energy without using VMC. When calculating the exact gradient, we find that the two paths converge to precisely the same accuracy ($\delta E/E_0\approx 10^{-5}$). However, when one uses VMC, hence noisy calculation of the gradient, we find that dynamical-$\chi$ leads to a significantly better accuracy. It is more than one order of magnitude better compared to the static optimization. It is also interesting that in the case of dynamical-$\chi$, the exact gradient optimization and the noisy one follow each other very closely. In contrast, for the static optimization the noisy gradient converges significantly slower compared to the exact gradient. In plain words, the naive intuition that it is easier to find one's bearings in a smaller space seems to be confirmed in this case.

Figure~\ref{fig:Breaking-of-optimisation1} shows a similar comparison of the optimization for the 2D model (where only the noisy gradient can be calculated). We find that dynamical-$\chi$ (black) gives a result of slightly better quality. But more importantly, the direct optimization fails completely for $\chi=7$, while no such problem occurs in the case of dynamical-$\chi$. Actually, we find that there is a tiny loss of $V$-score when we use dynamical-$\chi$ up to $\chi=7$. We surmise that the additional parameters do not improve the expressivity much while making the energy landscape harder to navigate.

We conclude that adopting an optimization strategy similar to the two-site DMRG algorithm is actually robust and efficient for the studied model and the \perceptrain ansatz, despite the ansatz being significantly different from a simple MPS.

\section{Technical matters}
\label{sec:technical}

In this last section, we give some technical details and highlight a few secondary findings. The model that we study presents a spontaneous symmetry breaking below a critical value of $h_x$. Section \ref{subsec:breaking} studies how our optimization process is affected by this spontaneous symmetry breaking. Rather surprisingly, we find that there exists a finite window of magnetic field where our ansatz is able to track both symmetry sectors simultaneously. Section \ref{subsec:GFMC} presents the GFMC method that we used to obtain the reference values for the energies used earlier for error calculations. GFMC is in principle an exact (in the sense of unbiased) method, a fact which we check explicitly. Section \ref{subsec:SR} looks at different variants of the gradient descent optimization. In contrast to neural network, we find that the optimization scheme is not crucial, although stochastic reconfiguration has slightly better performances. Section \ref{subsec:NonLin} looks at how the presence of non-linearity of the ansatz affects the stability of the optimization. 

\subsection{Observing spontaneous symmetry breaking within VMC}
\label{subsec:breaking}
We observe that the 2D model exhibits a quantum phase transition around $h_x \approx 2.4$ (for the studied size). For $h_x = 0$ the ground state is degenerate with two competing classical antiferromagnetic states. Switching all spins $s_i\to -s_i$ maps one of these antiferromagnetic states (tagged $\boldsymbol{s}_0$ below) onto the other (tagged $\boldsymbol{s}_V$ below). Hence, strictly speaking, the staggered magnetization vanishes at all $h_{x}$ due to the  $Z_2$ symmetry $s_i\to -s_i$. The classical way to break this symmetry is to add a small (staggered) Zeeman field to the Hamiltonian. 
 
Here, instead and mostly for the sake of curiosity, we do not break the $Z_2$ symmetry and study what happens during the optimization. To monitor the formation of a very large energy barrier between the two antiferromagnetic ground states, we consider a set of $V$ special configurations $\boldsymbol{s}_f$ that contain $f$ spins in one antiferromagnetic state and  $V-f$ spins in the other antiferromagnetic state with a domain wall in between, see the upper panel of Fig. \ref{fig:Wave-function-barrier} for illustration. Incrementally increasing $\boldsymbol{s}_f$ corresponds to a sequence of single spin-flips (allowed by the Hamiltonian) that move the domain wall. The highest energy states along this path  have $\sqrt{V}=10$ nearest neighbor interactions that are frustrated (the domain wall). They form the macroscopic barrier between the two $h_x=0$ ground states. Figure~\ref{fig:Wave-function-barrier}(a) shows $|\psi_{\boldsymbol{s}_f}/\psi_{\boldsymbol{s}_0}|^2$ versus $f$ while panel (b) shows $|\psi_{\boldsymbol{s}_{V}}/\psi_{\boldsymbol{s}_0}|^2$ (red) and $|\psi_{\boldsymbol{s}_{V/2}}/\psi_{\boldsymbol{s}_0}|^2$ (black) versus $h_x$. Very interestingly, we find that the optimization leads to three distinct regimes: for $h_x>2.4$, one is in the paramagnetic phase and the two symmetry sectors around $f=0$ and $f=V$ are equivalent $|\psi_{\boldsymbol{s}_{V}}/\psi_{\boldsymbol{s}_0}|^2 \approx 1$, meaning that one may "freely" go from one sector to the other $|\psi_{\boldsymbol{s}_{V/2}}/\psi_{\boldsymbol{s}_0}|^2 = O(1)$. For $h_x<2.22$ the optimization has spontaneously chosen one symmetry sector, $|\psi_{\boldsymbol{s}_{V}}/\psi_{\boldsymbol{s}_0}|^2\approx 0$ and $|\psi_{\boldsymbol{s}_{V/2}}/\psi_{\boldsymbol{s}_0}|^2\approx 0$. For $2.22<h_x<2.4$, we are in a mixed regime: the two symmetry sectors are well separated $|\psi_{\boldsymbol{s}_{V/2}}/\psi_{\boldsymbol{s}_0}|^2\approx 0$ but the wave function still captures \emph{both} symmetry sectors $|\psi_{\boldsymbol{s}_{V}}/\psi_{\boldsymbol{s}_0}|^2\approx 1$ [shaded gold region in Fig.~\ref{fig:Wave-function-barrier}b]. The fact that the wave function can track the two different $h_x=0$ ground states, although only for a finite window of magnetic fields, is quite promising for applications such as simulated quantum annealing. There, one would alternate optimization steps with a slow decrease of $h_x$, with the intent of finding the $h_x=0$ ground state, for a possibly highly frustrated Ising model.

\begin{figure}[t]
    \centering
        \includegraphics[width = 1.0\linewidth]{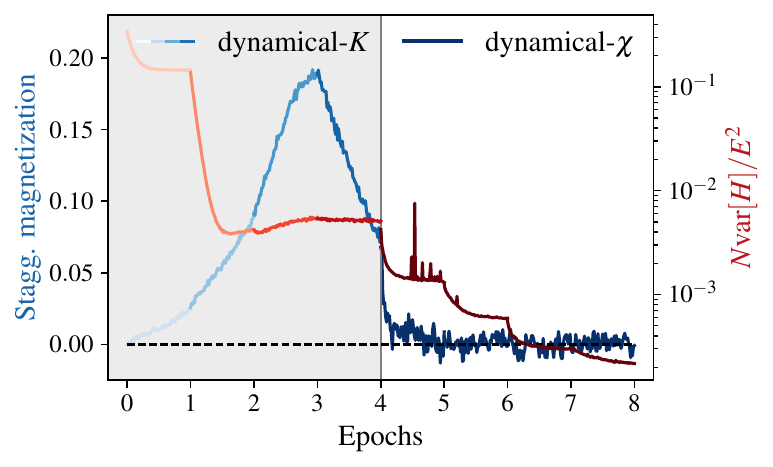}
    \caption{$V$-score (right axis) and staggered magnetization $M_z$ (left axis) as a function of the VMC optimization step for $h_x=2.3$ using dynamical-$K$ optimization in the SBS limit (epochs 1 to 4, $\alpha=10^{-4}$, $\chi=2$) followed by dynamical -$\chi$ optimization for the \perceptrain (epochs 5 to 8, $\alpha=10^{-3}$) limit.
    \label{fig:Magnetisation-along-VMC-optimisation}}
\end{figure}

The direct consequence of the existence of these three regimes is an hysteresis in the $M_z(h_x)$ curve. Figure~\ref{fig:Wave-function-barrier}(c) shows the result of two calculations where all the walkers have been initialized in the $\boldsymbol{s}_0$ configuration (red) or in the $\boldsymbol{s}_{V}$ (yellow) configuration. After thermalization of the Metropolis Markov chain, the walkers remain trapped around the original configuration for $2.22<h_x<2.4$. In this regime a fully random initialization of the walkers (with $50\%$ of the walkers in each sector) would lead to a vanishing staggered magnetization (which is mathematically correct but physically wrong). This case is illustrated in Fig.~\ref{fig:Magnetisation-along-VMC-optimisation}. We observe that the magnetization initially grows but then eventually decreases. It is only when we start increasing the rank $\chi$ that the optimization is able to recover the two symmetry sectors. To break the symmetry explicitly, the main results of Fig.~\ref{fig:Physics} for $M_{z}$ have been obtained by initializing all the walkers in the $\boldsymbol{s}_0$ configuration.

\begin{figure}[b]
    \centering
        \includegraphics[width = 1.0\linewidth]{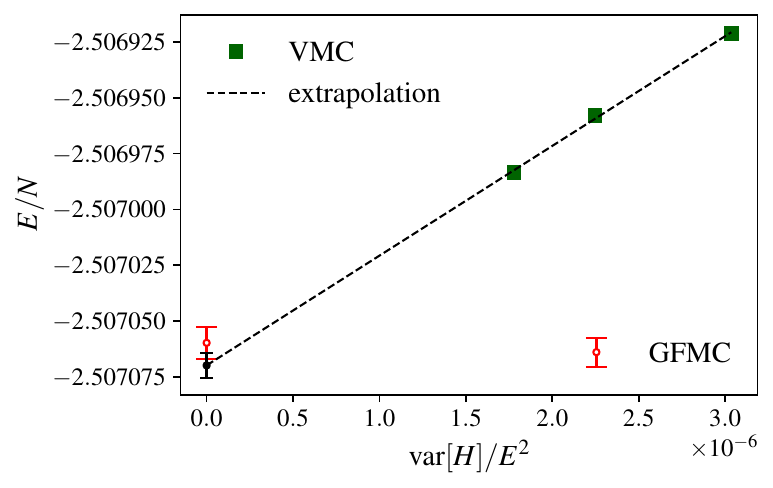}
    \caption{
    Linear zero variance extrapolation of the ground state energy of the 2D model at $h_{x}=2.3$, by considering the results at $\chi=4,5,6$. We plot the average (red dot), and the standard error (error bars) of the energy samples obtained by GFMC.
    \label{fig:VMC-extrapolation}}
\end{figure}

\begin{figure}[b]
    \centering
        \includegraphics[width = 1.0\linewidth]{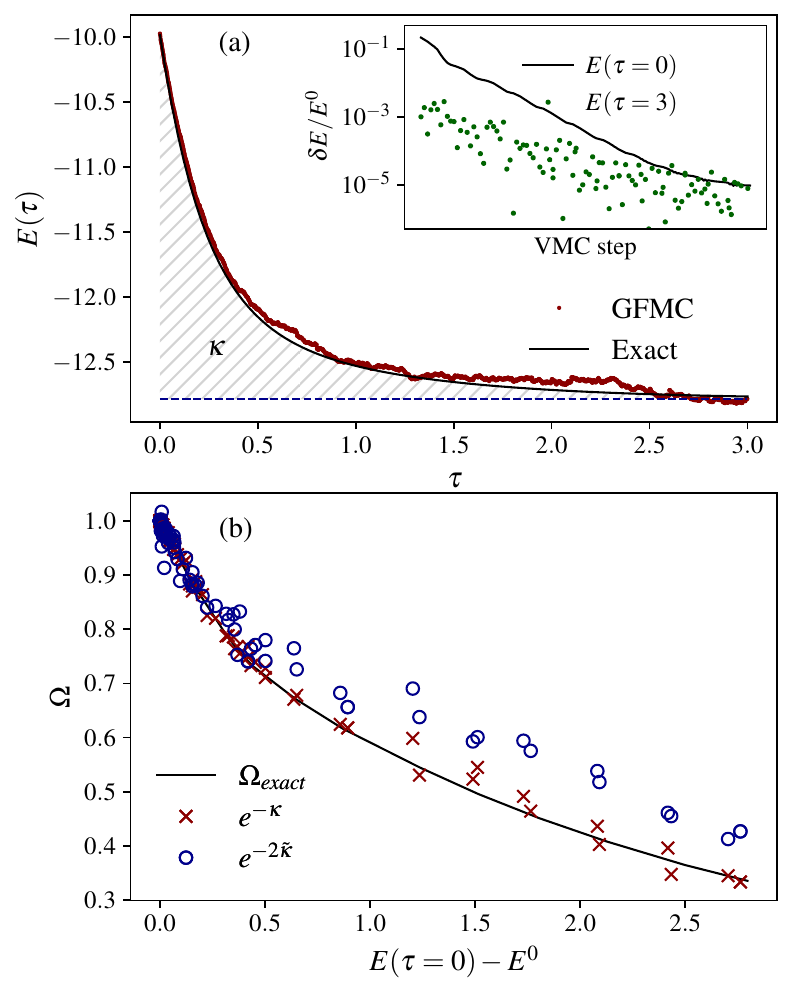}
    \caption{
    GFMC for the 1D model at $h_{x}=1$. The initial wave function is the ground state of $h_{x}=10$. (a) Exact curve $E(\tau)$ (black line) compared to the GFMC (red points). The overlap $\Omega$ from Eq.~\eqref{eq:overlap} is encoded in the area $\kappa$ (dashed region). (Inset) Variational energy [$E(\tau=0)$, black line] and associated GFMC energy [$E(\tau=3)$, green dots] versus VMC optimization step. For wave functions along the VMC optimization, the ground state energy $E(\tau=3)$ computed by GFMC (green) is compared to the initial guess' energy $E(\tau=0)$ (black line). (b) Overlap as a function of $E(\tau=0)$ for the data of the inset of panel (a). The estimators $e^{\kappa}$ (red crosses) from Eq.~\eqref{eq:GFMC-kappa} and $e^{2\tilde{\kappa}}$ (blue circles) from Eq.~\eqref{eq:GFMC-kappa-tilde} agree with the exact overlap (black line).
    \label{fig:GFMC}}
\end{figure}

\subsection{Benchmark with Green function Monte Carlo}
\label{subsec:GFMC}

To control the accuracy of our calculation, we use two different methods. The first, internal to VMC, is to extrapolate the curve $E(\text{var}[H]/E^{2})$ versus $\text{var}[H]\to0$. It can be shown that the scaling close to the ground state should be linear \cite{kashima_path-integral_2001}. Hence, we can use the energy variance pairs along the dynamical-$\chi$ or dynamical-$K$ optimization, for the linear extrapolation (see Fig.~\ref{fig:VMC-extrapolation} for an example). The difference between the extrapolation and the best VMC energy provides a good estimate of the error. This first method cannot distinguish the actual ground state from an excited state in case of e.g. a first order transition. The second method is to use the GFMC approach, a post VMC approach that projects the VMC ansatz onto the actual ground state of the system. GFMC is, in principle, exact. For the 1D model, we also have an independent exact calculation of the energy.

\subsubsection{GFMC in a nutshell}

GFMC is a lattice counterpart to the diffusion Monte Carlo approach \cite{reynolds_diffusion_1990}. Due to the presence of the lattice, GFMC can be formulated directly in the imaginary time continuum, without resorting to Trotterization or other discretization schemes. For completeness, we describe the method below and refer to the literature for the associated proofs \cite{trivedi_ground-state_1990,hida_crossover_1992,sorella_green_1998,sorella_green_2000,waintal_quantum_2006}. The VMC ansatz plays an important underlying role in GFMC and, as we shall see, condition the quality of the results. We describe here the version corresponding to a fixed sign of the Hamiltonian matrix elements, i.e. $H_{\boldsymbol{s}\boldsymbol{s'}} <0$ for all off-diagonal elements adapted to the transverse field Ising model. In this case, the sign of the ground state wave function is known for each configuration ($\psi_{\boldsymbol{s}}\ge 0$, absence of sign problem). When this property is absent, either the method becomes strongly inefficient or one uses the so-called fixed-node approximation to restore it at the cost of a (difficult to control) approximation.  

The goal of GFMC is to sample the state $|\psi (\tau)\rangle = e^{-H\tau} |\psi\rangle$ starting from the state  $|\psi\rangle$ represented by the VMC ansatz $\psi$. For $\tau$ large enough, $|\psi (\tau)\rangle$ converges to the actual ground state $|\psi^0\rangle$ of $H$ (in principle, for any choice of $\psi$ not orthogonal to $\psi^0$). The role of $\psi$ is both to be the starting point of the calculation but also to constrain the exploration of the configuration space. Hence, in the context of GFMC, $\psi$ is known as the "guiding" wave function. The primary output of GFMC is the curve $E(\tau)$ which starts at the VMC energy $E$ at $\tau=0$ and decreases toward $E^0$:
\begin{equation}
E(\tau) = \frac{\langle\psi| H e^{-H\tau} |\psi\rangle }{\langle\psi | e^{-H\tau} |\psi\rangle}.
\end{equation}
A secondary outcome of GFMC is the area $\kappa$ [see Fig.~\ref{fig:GFMC}(a)] under the curve $E(\tau)$,
\begin{equation}\label{eq:GFMC-kappa}
\kappa=\int \ [E(\tau)-E^0] d\tau,
\end{equation}
which is a very good metric for the quality of the variational ansatz. Indeed, it can be shown \cite{mora_variational_2007} that the overlap $\Omega$ between $|\psi\rangle$ and $|\psi^0\rangle$ (the so-called fidelity) is given by
\begin{equation}\label{eq:overlap}
\Omega \equiv \frac{|\langle\psi| \psi^{0}\rangle|^2}{\langle\psi|\psi\rangle  \langle\psi^0| \psi^0\rangle} = e^{-\kappa}.
\end{equation}
A good approximation of this area $\tilde \kappa$ is \cite{mora_variational_2007} (see Fig.~\ref{fig:GFMC}b),
\begin{equation}\label{eq:GFMC-kappa-tilde}
\tilde\kappa \equiv (E(0)-E^0)^2/2\text{var}[H] \approx \kappa.
\end{equation}
$\tilde \kappa$ is closely related to the $V$-score which provides a justification for the success of the later as a metric for the quality of the wave function.

The algorithm uses a set of $M$ walkers, each corresponding to a configuration ${\boldsymbol{s}}^\alpha (\tau)$ with $\alpha = 1...M$. Initially, the walkers must be distributed according to $|\psi|^2$. Each walker makes moves at discrete times distributed according to a Poisson statistics: the time $t$ that separates two moves is distributed according to 
\begin{equation}
p(t) = (H_{ {\boldsymbol{s}}^\alpha {\boldsymbol{s}}^\alpha } - e_{ {\boldsymbol{s}}^\alpha}  ) 
\exp\left[-( H_{ {\boldsymbol{s}}^\alpha {\boldsymbol{s}}^\alpha }  - e_{ {\boldsymbol{s}}^\alpha}  ) t\right],
\end{equation} 
which can be sampled directly. At each move, one of the spins is flipped. The choice of which of the $V$ spins is flipped is done according to 
\begin{equation}
p_{\boldsymbol{s'}} = \frac{\psi_{\boldsymbol{s'}}}{\sum_{\boldsymbol{s"}}\psi_{\boldsymbol{s"}}}
\end{equation}
where the sum $\sum_{\boldsymbol{s"}} $ extends to all the configurations connected to ${\boldsymbol{s}}^\alpha (\tau)$ by a single spin flip. At this point, the algorithm actually samples the VMC distribution $|\psi|^2$, it is a rejection free alternative to the Metropolis algorithm. To correct from the fact that one wants to sample from the actual ground state distribution $|\psi^0|^2$, one assigns a weight $w^\alpha$ to each walker. The weights are initialized to $w^\alpha(0)=1$. After a move, each weight is updated to 
\begin{equation}
w^\alpha(\tau+t)= w^\alpha(\tau) \exp( -e_{\boldsymbol{s}^\alpha(\tau)} t).
\end{equation}
The estimator of the energy is the weighted average,
\begin{equation}
E(\tau) \approx \frac{\sum_\alpha e_{ {\boldsymbol{s}}^\alpha(\tau)} w^\alpha (\tau) }
{\sum_\alpha w^\alpha(\tau)}.
\end{equation}
The above algorithm is exact (in the limit of large $M$) but inefficient because a handful of the walkers' weights will eventually become much larger than the others and dominate the average. Hence, this algorithm is completed with the so-called stochastic reconfiguration \cite{sorella_green_1998,sorella_green_2000} (same name, but not exactly the same mathematics as the one described in the first part of this manuscript). At fixed intervals, one performs the following "death and spawn" algorithm where some walkers are removed and others are duplicated while maintaining a fixed total number $M$ of walkers: one calculates $p_\alpha = M w^\alpha/\sum_{\alpha'} w^{\alpha'}$. Writing $p_\alpha$ as the sum of its floor and fractional part, $p_\alpha = \lfloor p_\alpha \rfloor + \delta p_\alpha$, then $\lfloor p_\alpha \rfloor$ copies of walker $\alpha$ are created (i.e a walker with $p_\alpha>1$ is certain to survive). One selects the remaining $M-\sum_\alpha \lfloor p_\alpha \rfloor$ walkers according to the probability $\delta p_\alpha$ (i.e. a walker with $p_\alpha<1$ may die). This stochastic reconfiguration introduces some correlations between different walkers while keeping the joint probability for $({\boldsymbol{s}}^\alpha , w^\alpha)$ unaffected. This concludes the algorithm.

\subsubsection{Benchmark for the one-dimensional model}

\begin{figure}[t]
    \centering
        \includegraphics[width = 1.0\linewidth]{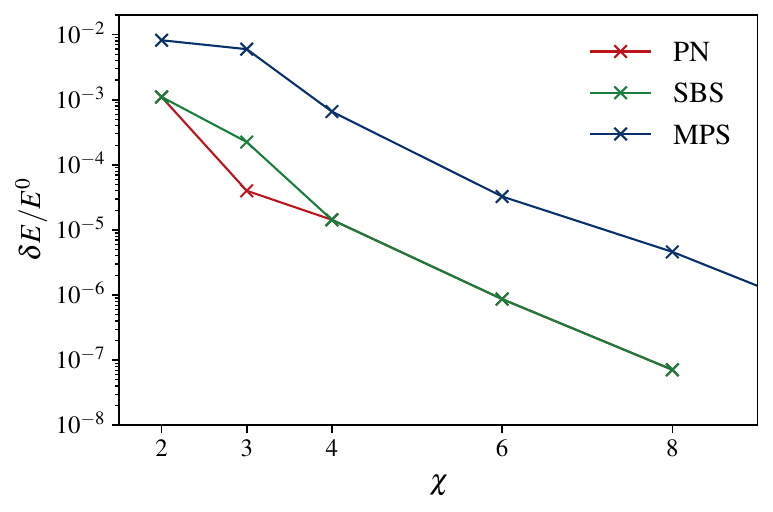}
    \caption{Relative energy error versus rank for the 1D model benchmark at $h_{x}=1$.  The \perceptrain with two identical child perceptrains (red), and the corresponding SBS limit (green) lead to at least one digit higher precision as MPS (blue) of the same rank $\chi$ even in this case particularly favorable to MPS. $d=5$. 
        \label{fig:Exact-VMC2}}
\end{figure}

The inset of Fig.~\ref{fig:GFMC}(a) shows an example of VMC optimization for the 1D model (black line). During the optimization, we perform many GFMC calculations (green points). We find consistently that GFMC improves the accuracy by one or two orders of magnitude until the VMC ansatz is so close to the ground state that one is dominated by statistical fluctuations. Main Fig.~\ref{fig:GFMC}(a) shows an example of the $E(\tau)$ using a very bad (almost constant) variational ansatz. We find that for this small system, GFMC is capable to converge to the ground-state energy nevertheless. For this simple case, we also calculated the exact curve $E(\tau)$ as a reference (black). The area $\kappa$ corresponds to the shaded area and is directly linked to $-\log \Omega$. Figure~\ref{fig:GFMC}(b) shows that a numerical integration of the $E(\tau)$ curve obtained in GFMC actually provides the correct fidelity $\Omega$. Also the proxy $\tilde\kappa$, which can be computed without GFMC, provides a correct (but more approximate) estimate of the overlap.

\begin{figure}[b]
    \centering
        \includegraphics[width = 1.0\linewidth]{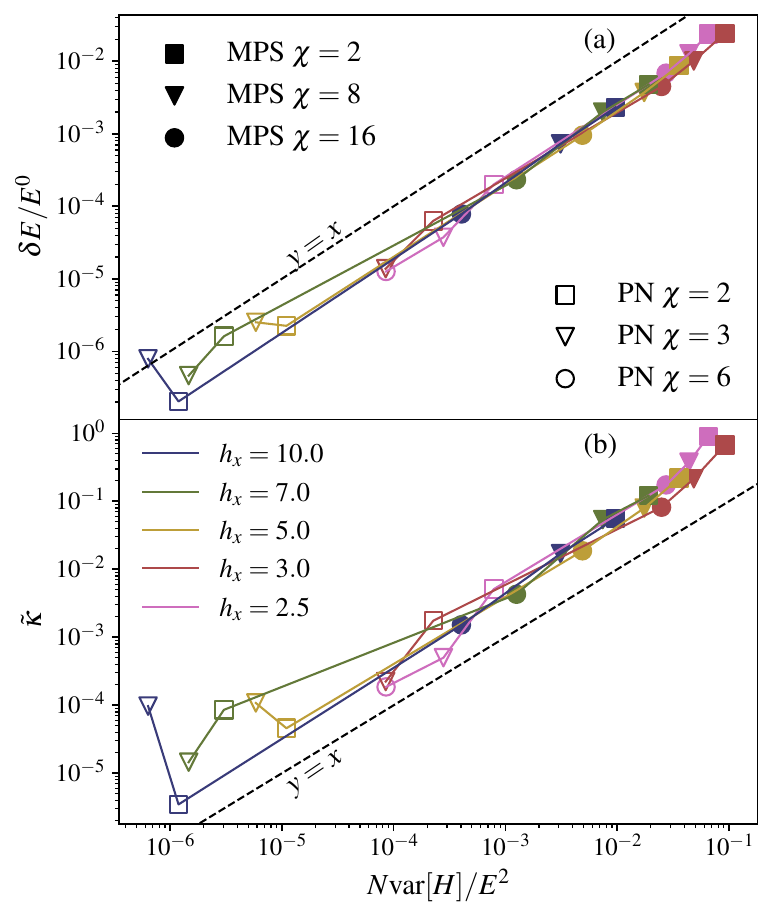}
    \caption{Best relative energy error (a) and $\tilde{\kappa}$ from Eq.~\eqref{eq:GFMC-kappa-tilde} (b) as a function of the $V$-score $N$var$[H]/E^{2}$ for different values of $h_{x}$. We see that the $V$-score roughly corresponds to (a) the number of digits in the energy and (b) the overlap with the ground state. ($\delta E / E^{0}$ is five times smaller than $V$-score, while $\tilde{\kappa}$ is five times bigger than $V$-score).
    \label{fig:Comparison-energ-and-kappa-vs-vscore}}
\end{figure}

\begin{figure}[t]
    \centering
        \includegraphics[width = 1.0\linewidth]{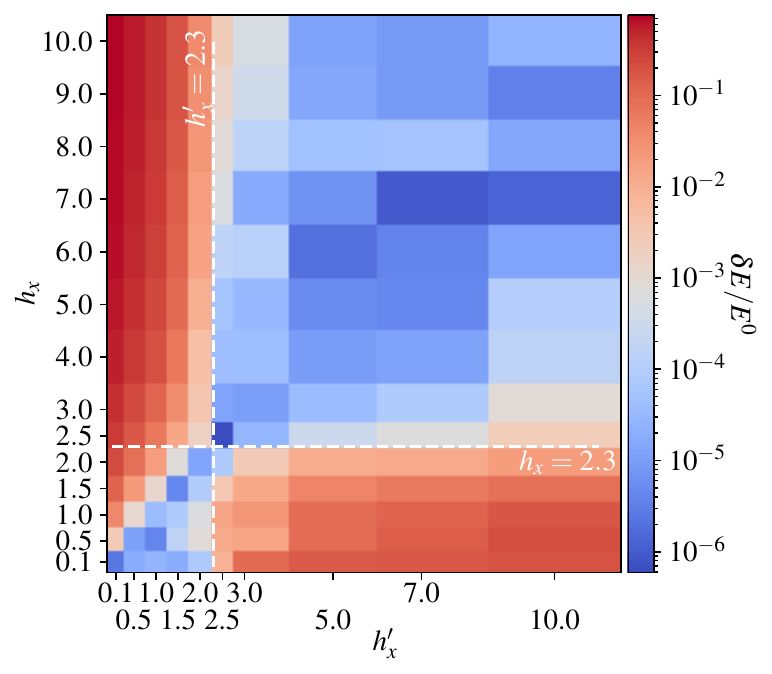}
    \caption{Role of the guiding wave function for the performance of GFMC for the 2D model. Relative energy error $\delta E/\delta E^{0} = |E(\tau=3) - E^{0}|/\delta E^{0}$ of the GFMC energy,  as a function of the transverse field $h_{x}$, and  a different transverse field $h_{x}'$ that was used for optimizing the guiding wave function (in a regular calculation, one uses $h_x=h_x'$). The result demonstrates that it is sufficient for the guiding wave function to be in the same phase (antiferromagnetic versus paramagnetic) for GFMC to converge to the correct energy (blue region) while large errors (red region) are obtained if $\psi$ and $\psi^0$ are in different phases.
    \label{fig:GFMC-energy-contour}}
\end{figure}

We used the 1D model also for comparing the performances of the MPS, the trivial \perceptrain consisting of two identical MPS, and the corresponding SBS limit. Our results for $h_x=1$ are summarized in Fig.~\ref{fig:Exact-VMC2}. We find that even for the 1D case, which is known to be very favorable to MPS, the \perceptrain ansatz consistently performs better than MPS by almost 2 digits. However, the \perceptrain ansatz is not better than the SBS limit (except at $\chi=3$, a point for which the meta-parameters $\alpha$, $\epsilon_1$, initialization, etc., have been optimized much more aggressively compared to the rest of the curve).

\begin{figure*}[t]
    \centering
        \includegraphics[width = 1.0\linewidth]{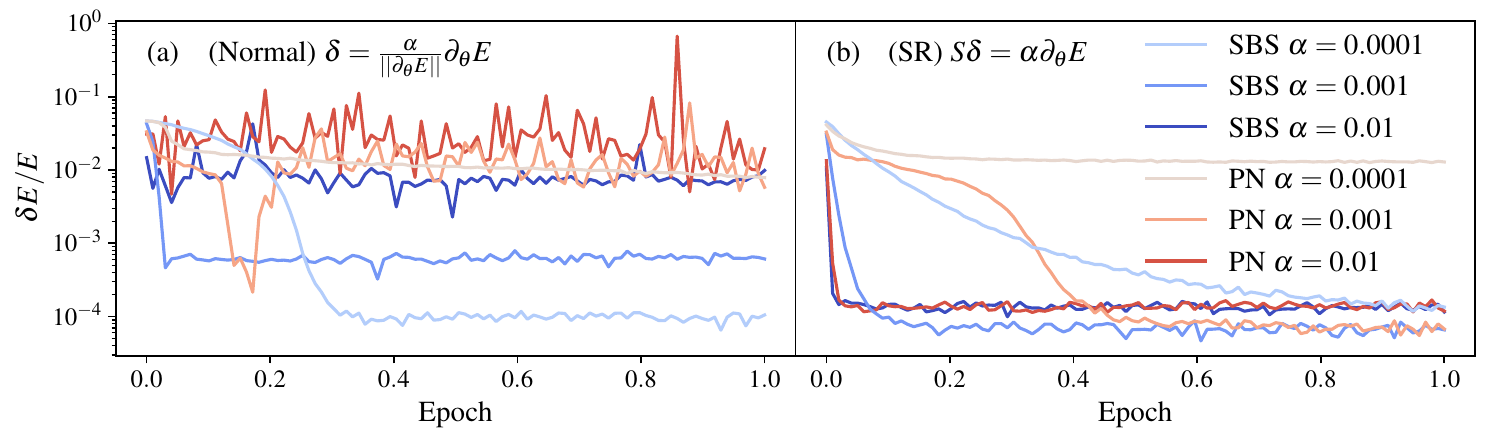}
    \caption{
    Relative energy error along the epoch for the 2D model at $h_{x}=3$ and $\chi=2$ \perceptrain (red lines), and the corresponding SBS limit (blue lines), for different learning rates $\alpha$. (a) With normal gradient descent, \perceptrain cannot be optimized, while SBS can be (final precision depends on $\alpha$) (b) With SR, both models can be optimized (\perceptrain remains more sensitive to $\alpha$).
    \label{fig:Optimisation-by-normal-vs-SR}}
\end{figure*}

\subsubsection{Benchmark for the two-dimensional model}

In this section, we present our reference calculations for the 2D model. We have performed various GFMC runs for different ans\"atze and values of $h_x$ and checked the consistency of the results. We have also, for completeness, compared the results with what could be obtained from a zero variance extrapolation \cite{kashima_path-integral_2001,nomura_helping_2021,chen_empowering_2024} (see Fig.~\ref{fig:VMC-extrapolation} for an example) and with the result of a regular MPS ansatz, as used in DMRG. The obtained values are the reference energies $E^{0}$ used throughout this manuscript to estimate errors. A sample of our most precise energies are presented in Table~\ref{tab:VMC-extrapolation}. At small values of $h_{x}=1,1.5$ the acceptance rate of the naïve Metropolis algorithm decreased leading to an increase of the auto-correlation time of the Markov process. In that regime, we switched to a rejection free sampling algorithm. In practice, the rejection free sampling is obtained by using GFMC in a mode where we force the weights of all walkers to remain equal to $1$ at all times.

In the process of elaborating this benchmark, we have made two important remarks. First, that the $V$-score is, at least for this model, a very good proxy for the relative precision on the energy [see Fig.~\ref{fig:Comparison-energ-and-kappa-vs-vscore}(a)] as well as the overlap $\tilde\kappa$  [see Fig.~\ref{fig:Comparison-energ-and-kappa-vs-vscore}(b)]. 

The second point is that the GFMC results are very robust, \emph{as long as} $\psi$ and $\psi^0$ are in the same phase (antiferromagnet versus paramagnet). This is illustrated very clearly in Fig.~\ref{fig:GFMC-energy-contour} which shows the error of the energy estimate as a function of $h_x$ (the actual $h_x$ of the Hamiltonian in the GFMC run) and $h_x'$ (the value of the transverse field that was used in the VMC optimization to construct the guiding wave function). In other words, in Fig.~\ref{fig:GFMC-energy-contour}, we have used "bad" guiding wave functions that have been optimized for a wrong value of the transverse field $h_x'\ne h_x$. The energies in Fig.~\ref{fig:GFMC-energy-contour} were averaged over the window $\tau\in[4, 6]$, and over $\tau\in[2, 6]$ on the diagonal. Random walks were performed using 32000 walkers and reconfiguration every $\Delta\tau= 0.1$.

We see that both $h_x$ and $h_x'$ must be in the same sector (i.e. same side with respect to the critical value $h_x=2.3$) for GFMC to succeed (blue region). In the antiferromagnetic sector, we observe an additional triangular structure which shows that the variational ansatz must be less localized than the true ground state; otherwise certain configurations are simply never visited.

\begin{figure}[b]
    \centering
        \includegraphics[width = 1.0\linewidth]{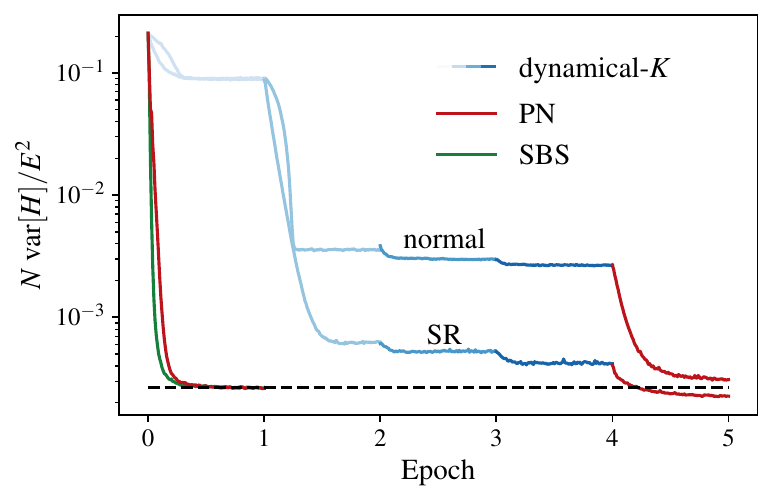}
    \caption{
     $V$-score along the VMC optimization of the 2D model at $h_x=3$ and fixed $\chi=2$. Dynamical-$K$ VMC optimization (blue) leads to slightly higher precision than the static VMC optimization, for \perceptrain (red) and SBS limit (green). The \perceptrain and SBS limit optimizations were all done using SR.
     \label{fig:Optimisation-by-increasing-vs-abrupt}}
\end{figure}

\begin{figure}[b]
    \centering
        \includegraphics[width = \linewidth]{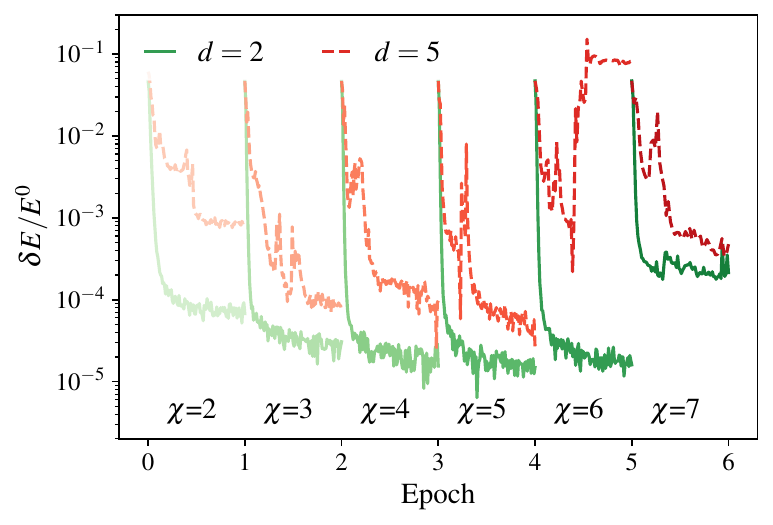}
    \caption{Relative energy error along the static VMC optimization for \perceptrain with $d=2$ (green) and $d=5$ (red), for 2D model at $h_{x}=3$. With more nonlinearities at higher $d$, the optimization becomes less stable. $\alpha=10^{-3}$.
   \label{fig:Breaking-of-optimisation}}
\end{figure}

\subsection{Stochastic reconfiguration versus normalized gradient descent}
\label{subsec:SR}

Figure~\ref{fig:Optimisation-by-normal-vs-SR} shows the result of the optimization for the \perceptrain ansatz (red) and in the SBS limit (blue) without SR (left panel) and with SR (right panel). We find that the optimization in the SBS limit is much more robust than for the \perceptrain ansatz, as a good accuracy can be reached even using the normalized gradient descent. However, SR is consistently more accurate and does not require a fine tuning of the $\alpha$ parameter. We recall that, in our scheme, there is no additional cost to using SR since we only optimize a small number of parameters at a given time (in contrast to usual gradient descent where all parameters are optimized simultaneously). In the same spirit, Fig.~\ref{fig:Optimisation-by-increasing-vs-abrupt} shows different optimization paths using/not using SR and using/not using dynamical-$K$ optimization. The optimization is globally very robust to these changes.

\subsection{Robustness of the optimization with respect to the presence of nonlinearities in the ansatz}
\label{subsec:NonLin}
We have found empirically that increasing the value of $d$, i.e. making the perceptrain more non-linear, makes the optimization less stable. This is illustrated in Fig.~\ref{fig:Breaking-of-optimisation}, which contrasts $d=2$ with $d=5$: the $d=5$ results are consistently less precise than $d=2$ and the optimization becomes unstable at a smaller rank ($\chi=6$ instead of $\chi=7$). In principle the $d=5$ ansatz \emph{contains} the full subspace of the $d=2$ one. Hence we conclude that the poor results obtained for $d=5$ are associated with an energy landscape that becomes harder to navigate. We attribute this difficulty to optimize to an inefficiency of the additional parameters introduced: they do not provide significant extra expressivity but make the energy landscape harder to navigate. Altogether, it seems that the SBS limit with a large value of $K$ is a good compromise.

We also observed that in the limit of small $h_{x}$, where the ground state converges to a product state and should therefore be well approximated by a low rank MPS ($\chi=1$ is exact in the limit of $h_x=0$), the optimization becomes unstable earlier. We find that when the rank is increased above $\chi=2$, both MPS and SBS errors become large in a manner reminiscent of Fig.~\ref{fig:Breaking-of-optimisation} (not shown). Again, we attribute this instability to the energy landscape being harder to navigate.

\section{Conclusions}
\label{sec:Conclusions}

In this work, we have explored how a simple network of "perceptrains," i.e. a neural network whose neuron units have been replaced by matrix product states, can be used to construct a variational ansatz for the 2D transverse field Ising model. We have learned a number of important lessons:
\begin{itemize}
\item First, the ansatz is accurate (typically five digits) for the entire range of transverse fields using ranks that would be considered as "tiny" in the tensor network community. This is very promising.
\item Second, a local optimization strategy, reminiscent of the one of the DMRG algorithm, provides a robust path to the ground state. This is also very important as variational approaches can be easily plagued by the difficulty to optimize (local minimum, barren plateaus...)
\item Third, the ability to change the number of parameters dynamically during the optimization is very relevant. It serves both to stabilize the optimization in presence of the (VMC induced) noise but also to use an optimal number of variational parameters. Overall, we obtain a very stable optimization.
\item Fourth, the main strength of our ansatz comes from the children MPS, not the master perceptrain. Indeed, the SBS limit, with the master MPS at $\chi=1$, and very limited expressive power, reaches the accuracy similar to the one of the full ansatz. Each of these child MPS can be seen as a complex ``feature'' capturing the correlations along certain directions (for example horizontal or vertical).
\end{itemize}
We believe that these features make perceptrain networks very strong candidates to challenge (or be combined with) regular neural networks for many-body ansatz and perhaps be useful in other contexts as well. 

On the physics side, we could essentially solve the two dimensional transverse field Ising model on a square lattice. The model is relevant for describing lattices of atoms in a superposition of ground and Rydberg states. Since the Rydberg platform has been proffered as a solver for discrete optimization problems, we can use our results to put bounds on the performance needed to reach the so-called quantum advantage.

Future works will explore different directions. First, our technique is directly applicable to more complex models of quantum magnetism or fermionic models, such as the Hubbard model. This requires a straightforward generalization from real to complex wave functions which, our preliminary results indicate, should not be problematic. The performance of the present approach should certainly be explored in these models. Second, further technical aspects can be improved: local optimization should allow one to use tailored optimization to speed up convergence; the choice of the orderings could be optimized and integrated in the optimization algorithm; one could use more expressive tensor networks, such as tensor trees \cite{shi_classical_2006,cheng_tree_2019} at no additional cost; one could build deeper and wider perceptrain networks. Third, the ansatz could also be used for studying the actual dynamics of the problem, within the time-dependent variational principle formalism.

\bigskip

\begin{acknowledgments}
The authors would also like to thank Sidhartha Dash for bringing to our attention the works \cite{schuch_strings_2008,sfondrini_simulating_2010,glasser_neural-network_2018}. MS would like to thank Antoine Georges' group at College de France for hosting him during his stay in Paris. MS and XW acknowledge the Plan France 2030 ANR-22-PETQ-0007 “EPIQ”, the PEPR “EQUBITFLY”, the ANR “DADI” and the CEA-FZJ French-German project AIDAS for the funding.
\end{acknowledgments}

\bibliography{Citations.bib}

\providecommand{\noopsort}[1]{}\providecommand{\singleletter}[1]{#1}%
\begin{thebibliography}{53}%
\makeatletter
\providecommand \@ifxundefined [1]{%
 \@ifx{#1\undefined}
}%
\providecommand \@ifnum [1]{%
 \ifnum #1\expandafter \@firstoftwo
 \else \expandafter \@secondoftwo
 \fi
}%
\providecommand \@ifx [1]{%
 \ifx #1\expandafter \@firstoftwo
 \else \expandafter \@secondoftwo
 \fi
}%
\providecommand \natexlab [1]{#1}%
\providecommand \enquote  [1]{``#1''}%
\providecommand \bibnamefont  [1]{#1}%
\providecommand \bibfnamefont [1]{#1}%
\providecommand \citenamefont [1]{#1}%
\providecommand \href@noop [0]{\@secondoftwo}%
\providecommand \href [0]{\begingroup \@sanitize@url \@href}%
\providecommand \@href[1]{\@@startlink{#1}\@@href}%
\providecommand \@@href[1]{\endgroup#1\@@endlink}%
\providecommand \@sanitize@url [0]{\catcode `\\12\catcode `\$12\catcode
  `\&12\catcode `\#12\catcode `\^12\catcode `\_12\catcode `\%12\relax}%
\providecommand \@@startlink[1]{}%
\providecommand \@@endlink[0]{}%
\providecommand \url  [0]{\begingroup\@sanitize@url \@url }%
\providecommand \@url [1]{\endgroup\@href {#1}{\urlprefix }}%
\providecommand \urlprefix  [0]{URL }%
\providecommand \Eprint [0]{\href }%
\providecommand \doibase [0]{https://doi.org/}%
\providecommand \selectlanguage [0]{\@gobble}%
\providecommand \bibinfo  [0]{\@secondoftwo}%
\providecommand \bibfield  [0]{\@secondoftwo}%
\providecommand \translation [1]{[#1]}%
\providecommand \BibitemOpen [0]{}%
\providecommand \bibitemStop [0]{}%
\providecommand \bibitemNoStop [0]{.\EOS\space}%
\providecommand \EOS [0]{\spacefactor3000\relax}%
\providecommand \BibitemShut  [1]{\csname bibitem#1\endcsname}%
\let\auto@bib@innerbib\@empty
\bibitem [{\citenamefont {Vaswani}\ \emph {et~al.}(2023)\citenamefont
  {Vaswani}, \citenamefont {Shazeer}, \citenamefont {Parmar}, \citenamefont
  {Uszkoreit}, \citenamefont {Jones}, \citenamefont {Gomez}, \citenamefont
  {Kaiser},\ and\ \citenamefont {Polosukhin}}]{vaswani_attention_2023}%
  \BibitemOpen
  \bibfield  {author} {\bibinfo {author} {\bibfnamefont {A.}~\bibnamefont
  {Vaswani}}, \bibinfo {author} {\bibfnamefont {N.}~\bibnamefont {Shazeer}},
  \bibinfo {author} {\bibfnamefont {N.}~\bibnamefont {Parmar}}, \bibinfo
  {author} {\bibfnamefont {J.}~\bibnamefont {Uszkoreit}}, \bibinfo {author}
  {\bibfnamefont {L.}~\bibnamefont {Jones}}, \bibinfo {author} {\bibfnamefont
  {A.~N.}\ \bibnamefont {Gomez}}, \bibinfo {author} {\bibfnamefont
  {L.}~\bibnamefont {Kaiser}},\ and\ \bibinfo {author} {\bibfnamefont
  {I.}~\bibnamefont {Polosukhin}},\ }\href
  {https://doi.org/10.48550/arXiv.1706.03762} {\bibinfo {title} {Attention is
  all you need}} (\bibinfo {year} {2023}),\ \bibinfo {note}
  {arXiv:1706.03762}\BibitemShut {NoStop}%
\bibitem [{\citenamefont {Nunez~Fernandez}\ \emph {et~al.}(2025)\citenamefont
  {Nunez~Fernandez}, \citenamefont {Ritter}, \citenamefont {Jeannin},
  \citenamefont {Li}, \citenamefont {Kloss}, \citenamefont {Louvet},
  \citenamefont {Terasaki}, \citenamefont {Parcollet}, \citenamefont {von
  Delft}, \citenamefont {Shinaoka},\ and\ \citenamefont
  {Waintal}}]{nunez_fernandez_learning_2025}%
  \BibitemOpen
  \bibfield  {author} {\bibinfo {author} {\bibfnamefont {Y.}~\bibnamefont
  {Nunez~Fernandez}}, \bibinfo {author} {\bibfnamefont {M.~K.}\ \bibnamefont
  {Ritter}}, \bibinfo {author} {\bibfnamefont {M.}~\bibnamefont {Jeannin}},
  \bibinfo {author} {\bibfnamefont {J.-W.}\ \bibnamefont {Li}}, \bibinfo
  {author} {\bibfnamefont {T.}~\bibnamefont {Kloss}}, \bibinfo {author}
  {\bibfnamefont {T.}~\bibnamefont {Louvet}}, \bibinfo {author} {\bibfnamefont
  {S.}~\bibnamefont {Terasaki}}, \bibinfo {author} {\bibfnamefont
  {O.}~\bibnamefont {Parcollet}}, \bibinfo {author} {\bibfnamefont
  {J.}~\bibnamefont {von Delft}}, \bibinfo {author} {\bibfnamefont
  {H.}~\bibnamefont {Shinaoka}},\ and\ \bibinfo {author} {\bibfnamefont
  {X.}~\bibnamefont {Waintal}},\ }\bibfield  {title} {{\selectlanguage
  {english}\bibinfo {title} {Learning tensor networks with tensor cross
  interpolation: New algorithms and libraries}},\ }\href
  {https://doi.org/10.21468/SciPostPhys.18.3.104} {\bibfield  {journal}
  {\bibinfo  {journal} {SciPost Physics}\ }\textbf {\bibinfo {volume} {18}},\
  \bibinfo {pages} {104} (\bibinfo {year} {2025})}\BibitemShut {NoStop}%
\bibitem [{\citenamefont {White}(1992)}]{White1992a}%
  \BibitemOpen
  \bibfield  {author} {\bibinfo {author} {\bibfnamefont {S.~R.}\ \bibnamefont
  {White}},\ }\bibfield  {title} {\bibinfo {title} {Density matrix formulation
  for quantum renormalization groups},\ }\href
  {https://doi.org/10.1103/physrevlett.69.2863} {\bibfield  {journal} {\bibinfo
   {journal} {Physical Review Letters}\ }\textbf {\bibinfo {volume} {69}},\
  \bibinfo {pages} {2863} (\bibinfo {year} {1992})}\BibitemShut {NoStop}%
\bibitem [{\citenamefont {White}\ and\ \citenamefont
  {Noack}(1992)}]{White1992b}%
  \BibitemOpen
  \bibfield  {author} {\bibinfo {author} {\bibfnamefont {S.~R.}\ \bibnamefont
  {White}}\ and\ \bibinfo {author} {\bibfnamefont {R.~M.}\ \bibnamefont
  {Noack}},\ }\bibfield  {title} {\bibinfo {title} {Real-space quantum
  renormalization groups},\ }\href
  {https://doi.org/10.1103/physrevlett.68.3487} {\bibfield  {journal} {\bibinfo
   {journal} {Physical Review Letters}\ }\textbf {\bibinfo {volume} {68}},\
  \bibinfo {pages} {3487} (\bibinfo {year} {1992})}\BibitemShut {NoStop}%
\bibitem [{\citenamefont {Corboz}\ \emph {et~al.}(2010)\citenamefont {Corboz},
  \citenamefont {Or{\'u}s}, \citenamefont {Bauer},\ and\ \citenamefont
  {Vidal}}]{corboz_simulation_2010}%
  \BibitemOpen
  \bibfield  {author} {\bibinfo {author} {\bibfnamefont {P.}~\bibnamefont
  {Corboz}}, \bibinfo {author} {\bibfnamefont {R.}~\bibnamefont {Or{\'u}s}},
  \bibinfo {author} {\bibfnamefont {B.}~\bibnamefont {Bauer}},\ and\ \bibinfo
  {author} {\bibfnamefont {G.}~\bibnamefont {Vidal}},\ }\bibfield  {title}
  {\bibinfo {title} {Simulation of strongly correlated fermions in two spatial
  dimensions with fermionic projected entangled-pair states},\ }\href
  {https://doi.org/10.1103/PhysRevB.81.165104} {\bibfield  {journal} {\bibinfo
  {journal} {Phys. Rev. B}\ }\textbf {\bibinfo {volume} {81}},\ \bibinfo
  {pages} {165104} (\bibinfo {year} {2010})}\BibitemShut {NoStop}%
\bibitem [{\citenamefont {Xu}\ \emph {et~al.}(2024)\citenamefont {Xu},
  \citenamefont {Chung}, \citenamefont {Qin}, \citenamefont {Schollw{\"o}ck},
  \citenamefont {White},\ and\ \citenamefont {Zhang}}]{xu_coexistence_2024}%
  \BibitemOpen
  \bibfield  {author} {\bibinfo {author} {\bibfnamefont {H.}~\bibnamefont
  {Xu}}, \bibinfo {author} {\bibfnamefont {C.-M.}\ \bibnamefont {Chung}},
  \bibinfo {author} {\bibfnamefont {M.}~\bibnamefont {Qin}}, \bibinfo {author}
  {\bibfnamefont {U.}~\bibnamefont {Schollw{\"o}ck}}, \bibinfo {author}
  {\bibfnamefont {S.~R.}\ \bibnamefont {White}},\ and\ \bibinfo {author}
  {\bibfnamefont {S.}~\bibnamefont {Zhang}},\ }\bibfield  {title} {\bibinfo
  {title} {Coexistence of superconductivity with partially filled stripes in
  the {Hubbard} model},\ }\href {https://doi.org/10.1126/science.adh7691}
  {\bibfield  {journal} {\bibinfo  {journal} {Science}\ }\textbf {\bibinfo
  {volume} {384}},\ \bibinfo {pages} {adh7691} (\bibinfo {year}
  {2024})}\BibitemShut {NoStop}%
\bibitem [{\citenamefont {Qian}\ \emph {et~al.}(2024)\citenamefont {Qian},
  \citenamefont {Huang},\ and\ \citenamefont {Qin}}]{qian_clifford_2024}%
  \BibitemOpen
  \bibfield  {author} {\bibinfo {author} {\bibfnamefont {X.}~\bibnamefont
  {Qian}}, \bibinfo {author} {\bibfnamefont {J.}~\bibnamefont {Huang}},\ and\
  \bibinfo {author} {\bibfnamefont {M.}~\bibnamefont {Qin}},\ }\bibfield
  {title} {\bibinfo {title} {Augmenting density matrix renormalization group
  with clifford circuits},\ }\href
  {https://doi.org/10.1103/PhysRevLett.133.190402} {\bibfield  {journal}
  {\bibinfo  {journal} {Phys. Rev. Lett.}\ }\textbf {\bibinfo {volume} {133}},\
  \bibinfo {pages} {190402} (\bibinfo {year} {2024})}\BibitemShut {NoStop}%
\bibitem [{\citenamefont {Carleo}\ and\ \citenamefont
  {Troyer}(2017)}]{carleo_solving_2017}%
  \BibitemOpen
  \bibfield  {author} {\bibinfo {author} {\bibfnamefont {G.}~\bibnamefont
  {Carleo}}\ and\ \bibinfo {author} {\bibfnamefont {M.}~\bibnamefont
  {Troyer}},\ }\bibfield  {title} {\bibinfo {title} {Solving the quantum
  many-body problem with artificial neural networks},\ }\href
  {https://doi.org/10.1126/science.aag2302} {\bibfield  {journal} {\bibinfo
  {journal} {Science}\ }\textbf {\bibinfo {volume} {355}},\ \bibinfo {pages}
  {602} (\bibinfo {year} {2017})}\BibitemShut {NoStop}%
\bibitem [{\citenamefont {Nomura}\ \emph {et~al.}(2017)\citenamefont {Nomura},
  \citenamefont {Darmawan}, \citenamefont {Yamaji},\ and\ \citenamefont
  {Imada}}]{nomura_restricted_2017}%
  \BibitemOpen
  \bibfield  {author} {\bibinfo {author} {\bibfnamefont {Y.}~\bibnamefont
  {Nomura}}, \bibinfo {author} {\bibfnamefont {A.~S.}\ \bibnamefont
  {Darmawan}}, \bibinfo {author} {\bibfnamefont {Y.}~\bibnamefont {Yamaji}},\
  and\ \bibinfo {author} {\bibfnamefont {M.}~\bibnamefont {Imada}},\ }\bibfield
   {title} {\bibinfo {title} {Restricted {Boltzmann} machine learning for
  solving strongly correlated quantum systems},\ }\href
  {https://doi.org/10.1103/PhysRevB.96.205152} {\bibfield  {journal} {\bibinfo
  {journal} {Physical Review B}\ }\textbf {\bibinfo {volume} {96}},\ \bibinfo
  {pages} {205152} (\bibinfo {year} {2017})}\BibitemShut {NoStop}%
\bibitem [{\citenamefont {Vicentini}\ \emph {et~al.}(2019)\citenamefont
  {Vicentini}, \citenamefont {Biella}, \citenamefont {Regnault},\ and\
  \citenamefont {Ciuti}}]{vicentini_variational_2019}%
  \BibitemOpen
  \bibfield  {author} {\bibinfo {author} {\bibfnamefont {F.}~\bibnamefont
  {Vicentini}}, \bibinfo {author} {\bibfnamefont {A.}~\bibnamefont {Biella}},
  \bibinfo {author} {\bibfnamefont {N.}~\bibnamefont {Regnault}},\ and\
  \bibinfo {author} {\bibfnamefont {C.}~\bibnamefont {Ciuti}},\ }\bibfield
  {title} {\bibinfo {title} {Variational {Neural}-{Network} {Ansatz} for
  {Steady} {States} in {Open} {Quantum} {Systems}},\ }\href
  {https://doi.org/10.1103/PhysRevLett.122.250503} {\bibfield  {journal}
  {\bibinfo  {journal} {Physical Review Letters}\ }\textbf {\bibinfo {volume}
  {122}},\ \bibinfo {pages} {250503} (\bibinfo {year} {2019})}\BibitemShut
  {NoStop}%
\bibitem [{\citenamefont {Choo}\ \emph {et~al.}(2019)\citenamefont {Choo},
  \citenamefont {Neupert},\ and\ \citenamefont
  {Carleo}}]{choo_two-dimensional_2019}%
  \BibitemOpen
  \bibfield  {author} {\bibinfo {author} {\bibfnamefont {K.}~\bibnamefont
  {Choo}}, \bibinfo {author} {\bibfnamefont {T.}~\bibnamefont {Neupert}},\ and\
  \bibinfo {author} {\bibfnamefont {G.}~\bibnamefont {Carleo}},\ }\bibfield
  {title} {\bibinfo {title} {Two-dimensional frustrated ${J_1}$-${J_2}$ model
  studied with neural network quantum states},\ }\href
  {https://doi.org/10.1103/PhysRevB.100.125124} {\bibfield  {journal} {\bibinfo
   {journal} {Physical Review B}\ }\textbf {\bibinfo {volume} {100}},\ \bibinfo
  {pages} {125124} (\bibinfo {year} {2019})}\BibitemShut {NoStop}%
\bibitem [{\citenamefont {Hibat-Allah}\ \emph {et~al.}(2020)\citenamefont
  {Hibat-Allah}, \citenamefont {Ganahl}, \citenamefont {Hayward}, \citenamefont
  {Melko},\ and\ \citenamefont {Carrasquilla}}]{hibat-allah_recurrent_2020}%
  \BibitemOpen
  \bibfield  {author} {\bibinfo {author} {\bibfnamefont {M.}~\bibnamefont
  {Hibat-Allah}}, \bibinfo {author} {\bibfnamefont {M.}~\bibnamefont {Ganahl}},
  \bibinfo {author} {\bibfnamefont {L.~E.}\ \bibnamefont {Hayward}}, \bibinfo
  {author} {\bibfnamefont {R.~G.}\ \bibnamefont {Melko}},\ and\ \bibinfo
  {author} {\bibfnamefont {J.}~\bibnamefont {Carrasquilla}},\ }\bibfield
  {title} {\bibinfo {title} {Recurrent neural network wave functions},\ }\href
  {https://doi.org/10.1103/PhysRevResearch.2.023358} {\bibfield  {journal}
  {\bibinfo  {journal} {Physical Review Research}\ }\textbf {\bibinfo {volume}
  {2}},\ \bibinfo {pages} {023358} (\bibinfo {year} {2020})}\BibitemShut
  {NoStop}%
\bibitem [{\citenamefont {Lovato}\ \emph {et~al.}(2022)\citenamefont {Lovato},
  \citenamefont {Adams}, \citenamefont {Carleo},\ and\ \citenamefont
  {Rocco}}]{lovato_hidden-nucleons_2022}%
  \BibitemOpen
  \bibfield  {author} {\bibinfo {author} {\bibfnamefont {A.}~\bibnamefont
  {Lovato}}, \bibinfo {author} {\bibfnamefont {C.}~\bibnamefont {Adams}},
  \bibinfo {author} {\bibfnamefont {G.}~\bibnamefont {Carleo}},\ and\ \bibinfo
  {author} {\bibfnamefont {N.}~\bibnamefont {Rocco}},\ }\bibfield  {title}
  {\bibinfo {title} {Hidden-nucleons neural-network quantum states for the
  nuclear many-body problem},\ }\href
  {https://doi.org/10.1103/PhysRevResearch.4.043178} {\bibfield  {journal}
  {\bibinfo  {journal} {Physical Review Research}\ }\textbf {\bibinfo {volume}
  {4}},\ \bibinfo {pages} {043178} (\bibinfo {year} {2022})}\BibitemShut
  {NoStop}%
\bibitem [{\citenamefont {Robledo~Moreno}\ \emph {et~al.}(2022)\citenamefont
  {Robledo~Moreno}, \citenamefont {Carleo}, \citenamefont {Georges},\ and\
  \citenamefont {Stokes}}]{robledo_moreno_fermionic_2022}%
  \BibitemOpen
  \bibfield  {author} {\bibinfo {author} {\bibfnamefont {J.}~\bibnamefont
  {Robledo~Moreno}}, \bibinfo {author} {\bibfnamefont {G.}~\bibnamefont
  {Carleo}}, \bibinfo {author} {\bibfnamefont {A.}~\bibnamefont {Georges}},\
  and\ \bibinfo {author} {\bibfnamefont {J.}~\bibnamefont {Stokes}},\
  }\bibfield  {title} {\bibinfo {title} {Fermionic wave functions from
  neural-network constrained hidden states},\ }\href
  {https://doi.org/10.1073/pnas.2122059119} {\bibfield  {journal} {\bibinfo
  {journal} {Proceedings of the National Academy of Sciences}\ }\textbf
  {\bibinfo {volume} {119}},\ \bibinfo {pages} {e2122059119} (\bibinfo {year}
  {2022})}\BibitemShut {NoStop}%
\bibitem [{\citenamefont {Sharir}\ \emph {et~al.}(2022)\citenamefont {Sharir},
  \citenamefont {Shashua},\ and\ \citenamefont {Carleo}}]{sharir_neural_2022}%
  \BibitemOpen
  \bibfield  {author} {\bibinfo {author} {\bibfnamefont {O.}~\bibnamefont
  {Sharir}}, \bibinfo {author} {\bibfnamefont {A.}~\bibnamefont {Shashua}},\
  and\ \bibinfo {author} {\bibfnamefont {G.}~\bibnamefont {Carleo}},\
  }\bibfield  {title} {\bibinfo {title} {Neural tensor contractions and the
  expressive power of deep neural quantum states},\ }\href
  {https://doi.org/10.1103/PhysRevB.106.205136} {\bibfield  {journal} {\bibinfo
   {journal} {Phys. Rev. B}\ }\textbf {\bibinfo {volume} {106}},\ \bibinfo
  {pages} {205136} (\bibinfo {year} {2022})},\ \bibinfo {note} {publisher:
  American Physical Society}\BibitemShut {NoStop}%
\bibitem [{\citenamefont {Kim}\ \emph {et~al.}(2023)\citenamefont {Kim},
  \citenamefont {Pescia}, \citenamefont {Fore}, \citenamefont {Nys},
  \citenamefont {Carleo}, \citenamefont {Gandolfi}, \citenamefont
  {Hjorth-Jensen},\ and\ \citenamefont {Lovato}}]{kim_neural-network_2023}%
  \BibitemOpen
  \bibfield  {author} {\bibinfo {author} {\bibfnamefont {J.}~\bibnamefont
  {Kim}}, \bibinfo {author} {\bibfnamefont {G.}~\bibnamefont {Pescia}},
  \bibinfo {author} {\bibfnamefont {B.}~\bibnamefont {Fore}}, \bibinfo {author}
  {\bibfnamefont {J.}~\bibnamefont {Nys}}, \bibinfo {author} {\bibfnamefont
  {G.}~\bibnamefont {Carleo}}, \bibinfo {author} {\bibfnamefont
  {S.}~\bibnamefont {Gandolfi}}, \bibinfo {author} {\bibfnamefont
  {M.}~\bibnamefont {Hjorth-Jensen}},\ and\ \bibinfo {author} {\bibfnamefont
  {A.}~\bibnamefont {Lovato}},\ }\href
  {https://doi.org/10.48550/arXiv.2305.08831} {\bibinfo {title} {Neural-network
  quantum states for ultra-cold {Fermi} gases}} (\bibinfo {year} {2023}),\
  \bibinfo {note} {arXiv:2305.08831}\BibitemShut {NoStop}%
\bibitem [{\citenamefont {Chen}\ and\ \citenamefont
  {Heyl}(2024)}]{chen_empowering_2024}%
  \BibitemOpen
  \bibfield  {author} {\bibinfo {author} {\bibfnamefont {A.}~\bibnamefont
  {Chen}}\ and\ \bibinfo {author} {\bibfnamefont {M.}~\bibnamefont {Heyl}},\
  }\bibfield  {title} {{\selectlanguage {english}\bibinfo {title} {Empowering
  deep neural quantum states through efficient optimization}},\ }\href
  {https://doi.org/10.1038/s41567-024-02566-1} {\bibfield  {journal} {\bibinfo
  {journal} {Nature Physics}\ }\textbf {\bibinfo {volume} {20}},\ \bibinfo
  {pages} {1476} (\bibinfo {year} {2024})}\BibitemShut {NoStop}%
\bibitem [{\citenamefont {Dash}\ \emph {et~al.}(2025)\citenamefont {Dash},
  \citenamefont {Gravina}, \citenamefont {Vicentini}, \citenamefont {Ferrero},\
  and\ \citenamefont {Georges}}]{dash_efficiency_2025}%
  \BibitemOpen
  \bibfield  {author} {\bibinfo {author} {\bibfnamefont {S.}~\bibnamefont
  {Dash}}, \bibinfo {author} {\bibfnamefont {L.}~\bibnamefont {Gravina}},
  \bibinfo {author} {\bibfnamefont {F.}~\bibnamefont {Vicentini}}, \bibinfo
  {author} {\bibfnamefont {M.}~\bibnamefont {Ferrero}},\ and\ \bibinfo {author}
  {\bibfnamefont {A.}~\bibnamefont {Georges}},\ }\bibfield  {title}
  {{\selectlanguage {english}\bibinfo {title} {Efficiency of neural quantum
  states in light of the quantum geometric tensor}},\ }\href
  {https://doi.org/10.1038/s42005-025-02005-4} {\bibfield  {journal} {\bibinfo
  {journal} {Commun Phys}\ }\textbf {\bibinfo {volume} {8}},\ \bibinfo {pages}
  {1} (\bibinfo {year} {2025})}\BibitemShut {NoStop}%
\bibitem [{\citenamefont {Sandvik}\ and\ \citenamefont
  {Vidal}(2007)}]{sandvik_variational_2007}%
  \BibitemOpen
  \bibfield  {author} {\bibinfo {author} {\bibfnamefont {A.~W.}\ \bibnamefont
  {Sandvik}}\ and\ \bibinfo {author} {\bibfnamefont {G.}~\bibnamefont
  {Vidal}},\ }\bibfield  {title} {{\selectlanguage {english}\bibinfo {title}
  {Variational quantum {Monte} {Carlo} simulations with tensor-network
  states}},\ }\href {https://doi.org/10.1103/PhysRevLett.99.220602} {\bibfield
  {journal} {\bibinfo  {journal} {Physical Review Letters}\ }\textbf {\bibinfo
  {volume} {99}},\ \bibinfo {pages} {220602} (\bibinfo {year}
  {2007})}\BibitemShut {NoStop}%
\bibitem [{\citenamefont {Liu}\ \emph {et~al.}(2021)\citenamefont {Liu},
  \citenamefont {Huang}, \citenamefont {Gong},\ and\ \citenamefont
  {Gu}}]{liu_accurate_2021}%
  \BibitemOpen
  \bibfield  {author} {\bibinfo {author} {\bibfnamefont {W.-Y.}\ \bibnamefont
  {Liu}}, \bibinfo {author} {\bibfnamefont {Y.-Z.}\ \bibnamefont {Huang}},
  \bibinfo {author} {\bibfnamefont {S.-S.}\ \bibnamefont {Gong}},\ and\
  \bibinfo {author} {\bibfnamefont {Z.-C.}\ \bibnamefont {Gu}},\ }\bibfield
  {title} {\bibinfo {title} {Accurate simulation for finite projected entangled
  pair states in two dimensions},\ }\href
  {https://doi.org/10.1103/PhysRevB.103.235155} {\bibfield  {journal} {\bibinfo
   {journal} {Physical Review B}\ }\textbf {\bibinfo {volume} {103}},\ \bibinfo
  {pages} {235155} (\bibinfo {year} {2021})}\BibitemShut {NoStop}%
\bibitem [{\citenamefont {Naumann}\ \emph {et~al.}(2024)\citenamefont
  {Naumann}, \citenamefont {Weerda}, \citenamefont {Rizzi}, \citenamefont
  {Eisert},\ and\ \citenamefont {Schmoll}}]{naumann_rizzi_introduction_2024}%
  \BibitemOpen
  \bibfield  {author} {\bibinfo {author} {\bibfnamefont {J.}~\bibnamefont
  {Naumann}}, \bibinfo {author} {\bibfnamefont {E.~L.}\ \bibnamefont {Weerda}},
  \bibinfo {author} {\bibfnamefont {M.}~\bibnamefont {Rizzi}}, \bibinfo
  {author} {\bibfnamefont {J.}~\bibnamefont {Eisert}},\ and\ \bibinfo {author}
  {\bibfnamefont {P.}~\bibnamefont {Schmoll}},\ }\bibfield  {title}
  {{\selectlanguage {english}\bibinfo {title} {An introduction to infinite
  projected entangled-pair state methods for variational ground state
  simulations using automatic differentiation}},\ }\href
  {https://doi.org/10.21468/SciPostPhysLectNotes.86} {\bibfield  {journal}
  {\bibinfo  {journal} {SciPost Physics Lecture Notes}\ ,\ \bibinfo {pages}
  {086}} (\bibinfo {year} {2024})}\BibitemShut {NoStop}%
\bibitem [{\citenamefont {Wang}\ \emph {et~al.}(2023)\citenamefont {Wang},
  \citenamefont {Pan}, \citenamefont {Xu}, \citenamefont {Yang}, \citenamefont
  {Li},\ and\ \citenamefont {Cichocki}}]{wang_tensor_2023}%
  \BibitemOpen
  \bibfield  {author} {\bibinfo {author} {\bibfnamefont {M.}~\bibnamefont
  {Wang}}, \bibinfo {author} {\bibfnamefont {Y.}~\bibnamefont {Pan}}, \bibinfo
  {author} {\bibfnamefont {Z.}~\bibnamefont {Xu}}, \bibinfo {author}
  {\bibfnamefont {X.}~\bibnamefont {Yang}}, \bibinfo {author} {\bibfnamefont
  {G.}~\bibnamefont {Li}},\ and\ \bibinfo {author} {\bibfnamefont
  {A.}~\bibnamefont {Cichocki}},\ }\href {http://arxiv.org/abs/2302.09019}
  {\bibinfo {title} {Tensor {Networks} {Meet} {Neural} {Networks}: {A} {Survey}
  and {Future} {Perspectives}}} (\bibinfo {year} {2023}),\ \bibinfo {note}
  {arXiv:2302.09019}\BibitemShut {NoStop}%
\bibitem [{\citenamefont {Wu}\ \emph {et~al.}(2023)\citenamefont {Wu},
  \citenamefont {Rossi}, \citenamefont {Vicentini},\ and\ \citenamefont
  {Carleo}}]{wu_tensor-network_2023}%
  \BibitemOpen
  \bibfield  {author} {\bibinfo {author} {\bibfnamefont {D.}~\bibnamefont
  {Wu}}, \bibinfo {author} {\bibfnamefont {R.}~\bibnamefont {Rossi}}, \bibinfo
  {author} {\bibfnamefont {F.}~\bibnamefont {Vicentini}},\ and\ \bibinfo
  {author} {\bibfnamefont {G.}~\bibnamefont {Carleo}},\ }\bibfield  {title}
  {\bibinfo {title} {From tensor-network quantum states to tensorial recurrent
  neural networks},\ }\href {https://doi.org/10.1103/PhysRevResearch.5.L032001}
  {\bibfield  {journal} {\bibinfo  {journal} {Phys. Rev. Res.}\ }\textbf
  {\bibinfo {volume} {5}},\ \bibinfo {pages} {L032001} (\bibinfo {year}
  {2023})},\ \bibinfo {note} {publisher: American Physical Society}\BibitemShut
  {NoStop}%
\bibitem [{\citenamefont {Lami}\ \emph {et~al.}(2022)\citenamefont {Lami},
  \citenamefont {Carleo},\ and\ \citenamefont {Collura}}]{lami_matrix_2022}%
  \BibitemOpen
  \bibfield  {author} {\bibinfo {author} {\bibfnamefont {G.}~\bibnamefont
  {Lami}}, \bibinfo {author} {\bibfnamefont {G.}~\bibnamefont {Carleo}},\ and\
  \bibinfo {author} {\bibfnamefont {M.}~\bibnamefont {Collura}},\ }\bibfield
  {title} {\bibinfo {title} {Matrix product states with backflow
  correlations},\ }\href {https://doi.org/10.1103/PhysRevB.106.L081111}
  {\bibfield  {journal} {\bibinfo  {journal} {Phys. Rev. B}\ }\textbf {\bibinfo
  {volume} {106}},\ \bibinfo {pages} {L081111} (\bibinfo {year} {2022})},\
  \bibinfo {note} {publisher: American Physical Society}\BibitemShut {NoStop}%
\bibitem [{\citenamefont {Schuch}\ \emph {et~al.}(2008)\citenamefont {Schuch},
  \citenamefont {Wolf}, \citenamefont {Verstraete},\ and\ \citenamefont
  {Cirac}}]{schuch_strings_2008}%
  \BibitemOpen
  \bibfield  {author} {\bibinfo {author} {\bibfnamefont {N.}~\bibnamefont
  {Schuch}}, \bibinfo {author} {\bibfnamefont {M.~M.}\ \bibnamefont {Wolf}},
  \bibinfo {author} {\bibfnamefont {F.}~\bibnamefont {Verstraete}},\ and\
  \bibinfo {author} {\bibfnamefont {J.~I.}\ \bibnamefont {Cirac}},\ }\bibfield
  {title} {\bibinfo {title} {Strings, {Projected} {Entangled} {Pair} {States},
  and variational {Monte} {Carlo} methods},\ }\href
  {https://doi.org/10.1103/PhysRevLett.100.040501} {\bibfield  {journal}
  {\bibinfo  {journal} {Physical Review Letters}\ }\textbf {\bibinfo {volume}
  {100}},\ \bibinfo {pages} {040501} (\bibinfo {year} {2008})}\BibitemShut
  {NoStop}%
\bibitem [{\citenamefont {Sfondrini}\ \emph {et~al.}(2010)\citenamefont
  {Sfondrini}, \citenamefont {Cerrillo}, \citenamefont {Schuch},\ and\
  \citenamefont {Cirac}}]{sfondrini_simulating_2010}%
  \BibitemOpen
  \bibfield  {author} {\bibinfo {author} {\bibfnamefont {A.}~\bibnamefont
  {Sfondrini}}, \bibinfo {author} {\bibfnamefont {J.}~\bibnamefont {Cerrillo}},
  \bibinfo {author} {\bibfnamefont {N.}~\bibnamefont {Schuch}},\ and\ \bibinfo
  {author} {\bibfnamefont {J.~I.}\ \bibnamefont {Cirac}},\ }\bibfield  {title}
  {\bibinfo {title} {Simulating two- and three-dimensional frustrated quantum
  systems with string-bond states},\ }\href
  {https://doi.org/10.1103/PhysRevB.81.214426} {\bibfield  {journal} {\bibinfo
  {journal} {Physical Review B}\ }\textbf {\bibinfo {volume} {81}},\ \bibinfo
  {pages} {214426} (\bibinfo {year} {2010})}\BibitemShut {NoStop}%
\bibitem [{\citenamefont {Glasser}\ \emph {et~al.}(2018)\citenamefont
  {Glasser}, \citenamefont {Pancotti}, \citenamefont {August}, \citenamefont
  {Rodriguez},\ and\ \citenamefont {Cirac}}]{glasser_neural-network_2018}%
  \BibitemOpen
  \bibfield  {author} {\bibinfo {author} {\bibfnamefont {I.}~\bibnamefont
  {Glasser}}, \bibinfo {author} {\bibfnamefont {N.}~\bibnamefont {Pancotti}},
  \bibinfo {author} {\bibfnamefont {M.}~\bibnamefont {August}}, \bibinfo
  {author} {\bibfnamefont {I.~D.}\ \bibnamefont {Rodriguez}},\ and\ \bibinfo
  {author} {\bibfnamefont {J.~I.}\ \bibnamefont {Cirac}},\ }\bibfield  {title}
  {\bibinfo {title} {Neural-{Network} {Quantum} {States}, {String}-{Bond}
  {States}, and {Chiral} {Topological} {States}},\ }\href
  {https://doi.org/10.1103/PhysRevX.8.011006} {\bibfield  {journal} {\bibinfo
  {journal} {Physical Review X}\ }\textbf {\bibinfo {volume} {8}},\ \bibinfo
  {pages} {011006} (\bibinfo {year} {2018})}\BibitemShut {NoStop}%
\bibitem [{\citenamefont
  {Schollwoeck}(2005)}]{schollwoeck_density-matrix_2005}%
  \BibitemOpen
  \bibfield  {author} {\bibinfo {author} {\bibfnamefont {U.}~\bibnamefont
  {Schollwoeck}},\ }\bibfield  {title} {\bibinfo {title} {The density-matrix
  renormalization group},\ }\href {https://doi.org/10.1103/RevModPhys.77.259}
  {\bibfield  {journal} {\bibinfo  {journal} {Reviews of Modern Physics}\
  }\textbf {\bibinfo {volume} {77}},\ \bibinfo {pages} {259} (\bibinfo {year}
  {2005})}\BibitemShut {NoStop}%
\bibitem [{\citenamefont {Dolgov}\ and\ \citenamefont
  {Savostyanov}(2020)}]{dolgov_parallel_2020}%
  \BibitemOpen
  \bibfield  {author} {\bibinfo {author} {\bibfnamefont {S.}~\bibnamefont
  {Dolgov}}\ and\ \bibinfo {author} {\bibfnamefont {D.}~\bibnamefont
  {Savostyanov}},\ }\bibfield  {title} {\bibinfo {title} {Parallel cross
  interpolation for high-precision calculation of high-dimensional integrals},\
  }\href {https://doi.org/10.1016/j.cpc.2019.106869} {\bibfield  {journal}
  {\bibinfo  {journal} {Computer Physics Communications}\ }\textbf {\bibinfo
  {volume} {246}},\ \bibinfo {pages} {106869} (\bibinfo {year}
  {2020})}\BibitemShut {NoStop}%
\bibitem [{\citenamefont {Nunez~Fernandez}\ \emph {et~al.}(2022)\citenamefont
  {Nunez~Fernandez}, \citenamefont {Jeannin}, \citenamefont {Dumitrescu},
  \citenamefont {Kloss}, \citenamefont {Kaye}, \citenamefont {Parcollet},\ and\
  \citenamefont {Waintal}}]{nunez_fernandez_learning_2022}%
  \BibitemOpen
  \bibfield  {author} {\bibinfo {author} {\bibfnamefont {Y.}~\bibnamefont
  {Nunez~Fernandez}}, \bibinfo {author} {\bibfnamefont {M.}~\bibnamefont
  {Jeannin}}, \bibinfo {author} {\bibfnamefont {P.~T.}\ \bibnamefont
  {Dumitrescu}}, \bibinfo {author} {\bibfnamefont {T.}~\bibnamefont {Kloss}},
  \bibinfo {author} {\bibfnamefont {J.}~\bibnamefont {Kaye}}, \bibinfo {author}
  {\bibfnamefont {O.}~\bibnamefont {Parcollet}},\ and\ \bibinfo {author}
  {\bibfnamefont {X.}~\bibnamefont {Waintal}},\ }\bibfield  {title} {\bibinfo
  {title} {Learning {Feynman} {Diagrams} with {Tensor} {Trains}},\ }\href
  {https://doi.org/10.1103/PhysRevX.12.041018} {\bibfield  {journal} {\bibinfo
  {journal} {Physical Review X}\ }\textbf {\bibinfo {volume} {12}},\ \bibinfo
  {pages} {041018} (\bibinfo {year} {2022})}\BibitemShut {NoStop}%
\bibitem [{\citenamefont {Becca}\ and\ \citenamefont
  {Sorella}(2017)}]{becca_quantum_2017}%
  \BibitemOpen
  \bibfield  {author} {\bibinfo {author} {\bibfnamefont {F.}~\bibnamefont
  {Becca}}\ and\ \bibinfo {author} {\bibfnamefont {S.}~\bibnamefont
  {Sorella}},\ }\href {https://doi.org/10.1017/9781316417041} {\emph {\bibinfo
  {title} {Quantum Monte Carlo Approaches for Correlated Systems}}},\ \bibinfo
  {edition} {1st}\ ed.\ (\bibinfo  {publisher} {Cambridge University Press},\
  \bibinfo {year} {2017})\BibitemShut {NoStop}%
\bibitem [{\citenamefont {Louvet}\ \emph {et~al.}(2024)\citenamefont {Louvet},
  \citenamefont {Ayral},\ and\ \citenamefont
  {Waintal}}]{louvet_feasibility_2024}%
  \BibitemOpen
  \bibfield  {author} {\bibinfo {author} {\bibfnamefont {T.}~\bibnamefont
  {Louvet}}, \bibinfo {author} {\bibfnamefont {T.}~\bibnamefont {Ayral}},\ and\
  \bibinfo {author} {\bibfnamefont {X.}~\bibnamefont {Waintal}},\ }\href
  {https://doi.org/10.48550/arXiv.2306.02620} {\bibinfo {title} {On the
  feasibility of performing quantum chemistry calculations on quantum
  computers}} (\bibinfo {year} {2024}),\ \bibinfo {note}
  {arXiv:2306.02620}\BibitemShut {NoStop}%
\bibitem [{\citenamefont
  {Schollw{\"o}ck}(2011)}]{schollwock_density-matrix_2011}%
  \BibitemOpen
  \bibfield  {author} {\bibinfo {author} {\bibfnamefont {U.}~\bibnamefont
  {Schollw{\"o}ck}},\ }\bibfield  {title} {\bibinfo {title} {The density-matrix
  renormalization group in the age of matrix product states},\ }\href
  {https://doi.org/10.1016/j.aop.2010.09.012} {\bibfield  {journal} {\bibinfo
  {journal} {Annals of Physics}\ }\bibinfo {series} {January 2011 {Special}
  {Issue}},\ \textbf {\bibinfo {volume} {326}},\ \bibinfo {pages} {96}
  (\bibinfo {year} {2011})}\BibitemShut {NoStop}%
\bibitem [{\citenamefont {Sorella}(1998)}]{sorella_green_1998}%
  \BibitemOpen
  \bibfield  {author} {\bibinfo {author} {\bibfnamefont {S.}~\bibnamefont
  {Sorella}},\ }\bibfield  {title} {{\selectlanguage {english}\bibinfo {title}
  {Green {Function} {Monte} {Carlo} with {Stochastic} {Reconfiguration}}},\
  }\href {https://doi.org/10.1103/PhysRevLett.80.4558} {\bibfield  {journal}
  {\bibinfo  {journal} {Physical Review Letters}\ }\textbf {\bibinfo {volume}
  {80}},\ \bibinfo {pages} {4558} (\bibinfo {year} {1998})}\BibitemShut
  {NoStop}%
\bibitem [{\citenamefont {Sorella}\ and\ \citenamefont
  {Capriotti}(2000)}]{sorella_green_2000}%
  \BibitemOpen
  \bibfield  {author} {\bibinfo {author} {\bibfnamefont {S.}~\bibnamefont
  {Sorella}}\ and\ \bibinfo {author} {\bibfnamefont {L.}~\bibnamefont
  {Capriotti}},\ }\bibfield  {title} {{\selectlanguage {english}\bibinfo
  {title} {Green function monte carlo with stochastic reconfiguration: An
  effective remedy for the sign problem}},\ }\href
  {https://doi.org/10.1103/PhysRevB.61.2599} {\bibfield  {journal} {\bibinfo
  {journal} {Physical Review B}\ }\textbf {\bibinfo {volume} {61}},\ \bibinfo
  {pages} {2599} (\bibinfo {year} {2000})}\BibitemShut {NoStop}%
\bibitem [{\citenamefont {Casula}\ and\ \citenamefont
  {Sorella}(2003)}]{casula_geminal_2003}%
  \BibitemOpen
  \bibfield  {author} {\bibinfo {author} {\bibfnamefont {M.}~\bibnamefont
  {Casula}}\ and\ \bibinfo {author} {\bibfnamefont {S.}~\bibnamefont
  {Sorella}},\ }\bibfield  {title} {\bibinfo {title} {Geminal wavefunctions
  with {Jastrow} correlation: a first application to atoms},\ }\href
  {https://doi.org/10.1063/1.1604379} {\bibfield  {journal} {\bibinfo
  {journal} {The Journal of Chemical Physics}\ }\textbf {\bibinfo {volume}
  {119}},\ \bibinfo {pages} {6500} (\bibinfo {year} {2003})}\BibitemShut
  {NoStop}%
\bibitem [{\citenamefont {Filippi}\ and\ \citenamefont
  {Fahy}(2000)}]{filippi_optimal_2000}%
  \BibitemOpen
  \bibfield  {author} {\bibinfo {author} {\bibfnamefont {C.}~\bibnamefont
  {Filippi}}\ and\ \bibinfo {author} {\bibfnamefont {S.}~\bibnamefont {Fahy}},\
  }\bibfield  {title} {{\selectlanguage {english}\bibinfo {title} {Optimal
  orbitals from energy fluctuations in correlated wave functions}},\ }\href
  {https://doi.org/10.1063/1.480507} {\bibfield  {journal} {\bibinfo  {journal}
  {The Journal of Chemical Physics}\ }\textbf {\bibinfo {volume} {112}},\
  \bibinfo {pages} {3523} (\bibinfo {year} {2000})}\BibitemShut {NoStop}%
\bibitem [{\citenamefont {Roth}\ \emph {et~al.}(2023)\citenamefont {Roth},
  \citenamefont {Szabo},\ and\ \citenamefont
  {MacDonald}}]{roth_high-accuracy_2023}%
  \BibitemOpen
  \bibfield  {author} {\bibinfo {author} {\bibfnamefont {C.}~\bibnamefont
  {Roth}}, \bibinfo {author} {\bibfnamefont {A.}~\bibnamefont {Szabo}},\ and\
  \bibinfo {author} {\bibfnamefont {A.}~\bibnamefont {MacDonald}},\ }\bibfield
  {title} {\bibinfo {title} {High-accuracy variational {Monte} {Carlo} for
  frustrated magnets with deep neural networks},\ }\href
  {https://doi.org/10.1103/PhysRevB.108.054410} {\bibfield  {journal} {\bibinfo
   {journal} {Physical Review B}\ }\textbf {\bibinfo {volume} {108}},\ \bibinfo
  {pages} {054410} (\bibinfo {year} {2023})}\BibitemShut {NoStop}%
\bibitem [{\citenamefont {Wu}\ \emph {et~al.}(2024)\citenamefont {Wu},
  \citenamefont {Rossi}, \citenamefont {Vicentini}, \citenamefont
  {Astrakhantsev}, \citenamefont {Becca}, \citenamefont {Cao}, \citenamefont
  {Carrasquilla}, \citenamefont {Ferrari}, \citenamefont {Georges},
  \citenamefont {Hibat-Allah}, \citenamefont {Imada}, \citenamefont
  {L{\"a}uchli}, \citenamefont {Mazzola}, \citenamefont {Mezzacapo},
  \citenamefont {Millis}, \citenamefont {Robledo~Moreno}, \citenamefont
  {Neupert}, \citenamefont {Nomura}, \citenamefont {Nys}, \citenamefont
  {Parcollet}, \citenamefont {Pohle}, \citenamefont {Romero}, \citenamefont
  {Schmid}, \citenamefont {Silvester}, \citenamefont {Sorella}, \citenamefont
  {Tocchio}, \citenamefont {Wang}, \citenamefont {White}, \citenamefont
  {Wietek}, \citenamefont {Yang}, \citenamefont {Yang}, \citenamefont {Zhang},\
  and\ \citenamefont {Carleo}}]{wu_vscore_2024}%
  \BibitemOpen
  \bibfield  {author} {\bibinfo {author} {\bibfnamefont {D.}~\bibnamefont
  {Wu}}, \bibinfo {author} {\bibfnamefont {R.}~\bibnamefont {Rossi}}, \bibinfo
  {author} {\bibfnamefont {F.}~\bibnamefont {Vicentini}}, \bibinfo {author}
  {\bibfnamefont {N.}~\bibnamefont {Astrakhantsev}}, \bibinfo {author}
  {\bibfnamefont {F.}~\bibnamefont {Becca}}, \bibinfo {author} {\bibfnamefont
  {X.}~\bibnamefont {Cao}}, \bibinfo {author} {\bibfnamefont {J.}~\bibnamefont
  {Carrasquilla}}, \bibinfo {author} {\bibfnamefont {F.}~\bibnamefont
  {Ferrari}}, \bibinfo {author} {\bibfnamefont {A.}~\bibnamefont {Georges}},
  \bibinfo {author} {\bibfnamefont {M.}~\bibnamefont {Hibat-Allah}}, \bibinfo
  {author} {\bibfnamefont {M.}~\bibnamefont {Imada}}, \bibinfo {author}
  {\bibfnamefont {A.~M.}\ \bibnamefont {L{\"a}uchli}}, \bibinfo {author}
  {\bibfnamefont {G.}~\bibnamefont {Mazzola}}, \bibinfo {author} {\bibfnamefont
  {A.}~\bibnamefont {Mezzacapo}}, \bibinfo {author} {\bibfnamefont
  {A.}~\bibnamefont {Millis}}, \bibinfo {author} {\bibfnamefont
  {J.}~\bibnamefont {Robledo~Moreno}}, \bibinfo {author} {\bibfnamefont
  {T.}~\bibnamefont {Neupert}}, \bibinfo {author} {\bibfnamefont
  {Y.}~\bibnamefont {Nomura}}, \bibinfo {author} {\bibfnamefont
  {J.}~\bibnamefont {Nys}}, \bibinfo {author} {\bibfnamefont {O.}~\bibnamefont
  {Parcollet}}, \bibinfo {author} {\bibfnamefont {R.}~\bibnamefont {Pohle}},
  \bibinfo {author} {\bibfnamefont {I.}~\bibnamefont {Romero}}, \bibinfo
  {author} {\bibfnamefont {M.}~\bibnamefont {Schmid}}, \bibinfo {author}
  {\bibfnamefont {J.~M.}\ \bibnamefont {Silvester}}, \bibinfo {author}
  {\bibfnamefont {S.}~\bibnamefont {Sorella}}, \bibinfo {author} {\bibfnamefont
  {L.~F.}\ \bibnamefont {Tocchio}}, \bibinfo {author} {\bibfnamefont
  {L.}~\bibnamefont {Wang}}, \bibinfo {author} {\bibfnamefont {S.~R.}\
  \bibnamefont {White}}, \bibinfo {author} {\bibfnamefont {A.}~\bibnamefont
  {Wietek}}, \bibinfo {author} {\bibfnamefont {Q.}~\bibnamefont {Yang}},
  \bibinfo {author} {\bibfnamefont {Y.}~\bibnamefont {Yang}}, \bibinfo {author}
  {\bibfnamefont {S.}~\bibnamefont {Zhang}},\ and\ \bibinfo {author}
  {\bibfnamefont {G.}~\bibnamefont {Carleo}},\ }\bibfield  {title} {\bibinfo
  {title} {Variational benchmarks for quantum many-body problems},\ }\href
  {https://doi.org/10.1126/science.adg9774} {\bibfield  {journal} {\bibinfo
  {journal} {Science}\ }\textbf {\bibinfo {volume} {386}},\ \bibinfo {pages}
  {296} (\bibinfo {year} {2024})}\BibitemShut {NoStop}%
\bibitem [{\citenamefont {Samajdar}\ \emph {et~al.}(2020)\citenamefont
  {Samajdar}, \citenamefont {Ho}, \citenamefont {Pichler}, \citenamefont
  {Lukin},\ and\ \citenamefont {Sachdev}}]{samajdar_complex_2020}%
  \BibitemOpen
  \bibfield  {author} {\bibinfo {author} {\bibfnamefont {R.}~\bibnamefont
  {Samajdar}}, \bibinfo {author} {\bibfnamefont {W.~W.}\ \bibnamefont {Ho}},
  \bibinfo {author} {\bibfnamefont {H.}~\bibnamefont {Pichler}}, \bibinfo
  {author} {\bibfnamefont {M.~D.}\ \bibnamefont {Lukin}},\ and\ \bibinfo
  {author} {\bibfnamefont {S.}~\bibnamefont {Sachdev}},\ }\bibfield  {title}
  {\bibinfo {title} {Complex density wave orders and quantum phase transitions
  in a model of square-lattice rydberg atom arrays},\ }\href
  {https://doi.org/10.1103/PhysRevLett.124.103601} {\bibfield  {journal}
  {\bibinfo  {journal} {Physical Review Letters}\ }\textbf {\bibinfo {volume}
  {124}},\ \bibinfo {pages} {103601} (\bibinfo {year} {2020})}\BibitemShut
  {NoStop}%
\bibitem [{\citenamefont {Scholl}\ \emph {et~al.}(2021)\citenamefont {Scholl},
  \citenamefont {Schuler}, \citenamefont {Williams}, \citenamefont
  {Eberharter}, \citenamefont {Barredo}, \citenamefont {Schymik}, \citenamefont
  {Lienhard}, \citenamefont {Henry}, \citenamefont {Lang}, \citenamefont
  {Lahaye}, \citenamefont {L{\"a}uchli},\ and\ \citenamefont
  {Browaeys}}]{scholl_quantum_2021}%
  \BibitemOpen
  \bibfield  {author} {\bibinfo {author} {\bibfnamefont {P.}~\bibnamefont
  {Scholl}}, \bibinfo {author} {\bibfnamefont {M.}~\bibnamefont {Schuler}},
  \bibinfo {author} {\bibfnamefont {H.~J.}\ \bibnamefont {Williams}}, \bibinfo
  {author} {\bibfnamefont {A.~A.}\ \bibnamefont {Eberharter}}, \bibinfo
  {author} {\bibfnamefont {D.}~\bibnamefont {Barredo}}, \bibinfo {author}
  {\bibfnamefont {K.-N.}\ \bibnamefont {Schymik}}, \bibinfo {author}
  {\bibfnamefont {V.}~\bibnamefont {Lienhard}}, \bibinfo {author}
  {\bibfnamefont {L.-P.}\ \bibnamefont {Henry}}, \bibinfo {author}
  {\bibfnamefont {T.~C.}\ \bibnamefont {Lang}}, \bibinfo {author}
  {\bibfnamefont {T.}~\bibnamefont {Lahaye}}, \bibinfo {author} {\bibfnamefont
  {A.~M.}\ \bibnamefont {L{\"a}uchli}},\ and\ \bibinfo {author} {\bibfnamefont
  {A.}~\bibnamefont {Browaeys}},\ }\bibfield  {title} {{\selectlanguage
  {english}\bibinfo {title} {Quantum simulation of {2D} antiferromagnets with
  hundreds of {Rydberg} atoms}},\ }\href
  {https://doi.org/10.1038/s41586-021-03585-1} {\bibfield  {journal} {\bibinfo
  {journal} {Nature}\ }\textbf {\bibinfo {volume} {595}},\ \bibinfo {pages}
  {233} (\bibinfo {year} {2021})}\BibitemShut {NoStop}%
\bibitem [{\citenamefont {Ebadi}\ \emph {et~al.}(2021)\citenamefont {Ebadi},
  \citenamefont {Wang}, \citenamefont {Levine}, \citenamefont {Keesling},
  \citenamefont {Semeghini}, \citenamefont {Omran}, \citenamefont {Bluvstein},
  \citenamefont {Samajdar}, \citenamefont {Pichler}, \citenamefont {Ho},
  \citenamefont {Choi}, \citenamefont {Sachdev}, \citenamefont {Greiner},
  \citenamefont {Vuletic},\ and\ \citenamefont {Lukin}}]{ebadi_quantum_2021}%
  \BibitemOpen
  \bibfield  {author} {\bibinfo {author} {\bibfnamefont {S.}~\bibnamefont
  {Ebadi}}, \bibinfo {author} {\bibfnamefont {T.~T.}\ \bibnamefont {Wang}},
  \bibinfo {author} {\bibfnamefont {H.}~\bibnamefont {Levine}}, \bibinfo
  {author} {\bibfnamefont {A.}~\bibnamefont {Keesling}}, \bibinfo {author}
  {\bibfnamefont {G.}~\bibnamefont {Semeghini}}, \bibinfo {author}
  {\bibfnamefont {A.}~\bibnamefont {Omran}}, \bibinfo {author} {\bibfnamefont
  {D.}~\bibnamefont {Bluvstein}}, \bibinfo {author} {\bibfnamefont
  {R.}~\bibnamefont {Samajdar}}, \bibinfo {author} {\bibfnamefont
  {H.}~\bibnamefont {Pichler}}, \bibinfo {author} {\bibfnamefont {W.~W.}\
  \bibnamefont {Ho}}, \bibinfo {author} {\bibfnamefont {S.}~\bibnamefont
  {Choi}}, \bibinfo {author} {\bibfnamefont {S.}~\bibnamefont {Sachdev}},
  \bibinfo {author} {\bibfnamefont {M.}~\bibnamefont {Greiner}}, \bibinfo
  {author} {\bibfnamefont {V.}~\bibnamefont {Vuletic}},\ and\ \bibinfo {author}
  {\bibfnamefont {M.~D.}\ \bibnamefont {Lukin}},\ }\bibfield  {title} {\bibinfo
  {title} {Quantum {Phases} of {Matter} on a 256-{Atom} {Programmable}
  {Quantum} {Simulator}},\ }\href {https://doi.org/10.1038/s41586-021-03582-4}
  {\bibfield  {journal} {\bibinfo  {journal} {Nature}\ }\textbf {\bibinfo
  {volume} {595}},\ \bibinfo {pages} {227} (\bibinfo {year}
  {2021})}\BibitemShut {NoStop}%
\bibitem [{\citenamefont {Kalinowski}\ \emph {et~al.}(2022)\citenamefont
  {Kalinowski}, \citenamefont {Samajdar}, \citenamefont {Melko}, \citenamefont
  {Lukin}, \citenamefont {Sachdev},\ and\ \citenamefont
  {Choi}}]{kalinowski_bulk_2022}%
  \BibitemOpen
  \bibfield  {author} {\bibinfo {author} {\bibfnamefont {M.}~\bibnamefont
  {Kalinowski}}, \bibinfo {author} {\bibfnamefont {R.}~\bibnamefont
  {Samajdar}}, \bibinfo {author} {\bibfnamefont {R.~G.}\ \bibnamefont {Melko}},
  \bibinfo {author} {\bibfnamefont {M.~D.}\ \bibnamefont {Lukin}}, \bibinfo
  {author} {\bibfnamefont {S.}~\bibnamefont {Sachdev}},\ and\ \bibinfo {author}
  {\bibfnamefont {S.}~\bibnamefont {Choi}},\ }\bibfield  {title}
  {{\selectlanguage {english}\bibinfo {title} {Bulk and boundary quantum phase
  transitions in a square {Rydberg} atom array}},\ }\href
  {https://doi.org/10.1103/PhysRevB.105.174417} {\bibfield  {journal} {\bibinfo
   {journal} {Physical Review B}\ }\textbf {\bibinfo {volume} {105}},\ \bibinfo
  {pages} {174417} (\bibinfo {year} {2022})}\BibitemShut {NoStop}%
\bibitem [{\citenamefont {Kalz}\ \emph {et~al.}(2008)\citenamefont {Kalz},
  \citenamefont {Honecker}, \citenamefont {Fuchs},\ and\ \citenamefont
  {Pruschke}}]{kalz_phase_2008}%
  \BibitemOpen
  \bibfield  {author} {\bibinfo {author} {\bibfnamefont {A.}~\bibnamefont
  {Kalz}}, \bibinfo {author} {\bibfnamefont {A.}~\bibnamefont {Honecker}},
  \bibinfo {author} {\bibfnamefont {S.}~\bibnamefont {Fuchs}},\ and\ \bibinfo
  {author} {\bibfnamefont {T.}~\bibnamefont {Pruschke}},\ }\bibfield  {title}
  {\bibinfo {title} {Phase diagram of the {Ising} square lattice with competing
  interactions},\ }\href {https://doi.org/10.1140/epjb/e2008-00359-6}
  {\bibfield  {journal} {\bibinfo  {journal} {The European Physical Journal B}\
  }\textbf {\bibinfo {volume} {65}},\ \bibinfo {pages} {533} (\bibinfo {year}
  {2008})}\BibitemShut {NoStop}%
\bibitem [{\citenamefont {Kashima}\ and\ \citenamefont
  {Imada}(2001)}]{kashima_path-integral_2001}%
  \BibitemOpen
  \bibfield  {author} {\bibinfo {author} {\bibfnamefont {T.}~\bibnamefont
  {Kashima}}\ and\ \bibinfo {author} {\bibfnamefont {M.}~\bibnamefont
  {Imada}},\ }\bibfield  {title} {\bibinfo {title} {Path-{Integral}
  {Renormalization} {Group} {Method} for {Numerical} {Study} on {Ground}
  {States} of {Strongly} {Correlated} {Electronic} {Systems}},\ }\href
  {https://doi.org/10.1143/JPSJ.70.2287} {\bibfield  {journal} {\bibinfo
  {journal} {J. Phys. Soc. Jpn.}\ }\textbf {\bibinfo {volume} {70}},\ \bibinfo
  {pages} {2287} (\bibinfo {year} {2001})}\BibitemShut {NoStop}%
\bibitem [{\citenamefont {Reynolds}\ \emph {et~al.}(1990)\citenamefont
  {Reynolds}, \citenamefont {Tobochnik},\ and\ \citenamefont
  {Gould}}]{reynolds_diffusion_1990}%
  \BibitemOpen
  \bibfield  {author} {\bibinfo {author} {\bibfnamefont {P.~J.}\ \bibnamefont
  {Reynolds}}, \bibinfo {author} {\bibfnamefont {J.}~\bibnamefont
  {Tobochnik}},\ and\ \bibinfo {author} {\bibfnamefont {H.}~\bibnamefont
  {Gould}},\ }\bibfield  {title} {\bibinfo {title} {Diffusion {Quantum} {Monte}
  {Carlo}},\ }\href {https://doi.org/10.1063/1.4822960} {\bibfield  {journal}
  {\bibinfo  {journal} {Computer in Physics}\ }\textbf {\bibinfo {volume}
  {4}},\ \bibinfo {pages} {662} (\bibinfo {year} {1990})}\BibitemShut {NoStop}%
\bibitem [{\citenamefont {Trivedi}\ and\ \citenamefont
  {Ceperley}(1990)}]{trivedi_ground-state_1990}%
  \BibitemOpen
  \bibfield  {author} {\bibinfo {author} {\bibfnamefont {N.}~\bibnamefont
  {Trivedi}}\ and\ \bibinfo {author} {\bibfnamefont {D.~M.}\ \bibnamefont
  {Ceperley}},\ }\bibfield  {title} {{\selectlanguage {english}\bibinfo {title}
  {Ground-state correlations of quantum antiferromagnets: {A} {Green}-function
  {Monte} {Carlo} study}},\ }\href {https://doi.org/10.1103/PhysRevB.41.4552}
  {\bibfield  {journal} {\bibinfo  {journal} {Physical Review B}\ }\textbf
  {\bibinfo {volume} {41}},\ \bibinfo {pages} {4552} (\bibinfo {year}
  {1990})}\BibitemShut {NoStop}%
\bibitem [{\citenamefont {Hida}(1992)}]{hida_crossover_1992}%
  \BibitemOpen
  \bibfield  {author} {\bibinfo {author} {\bibfnamefont {K.}~\bibnamefont
  {Hida}},\ }\bibfield  {title} {\bibinfo {title} {Crossover between the
  {Haldane}-gap phase and the dimer phase in the spin-1/2 alternating
  {Heisenberg} chain},\ }\href {https://doi.org/10.1103/PhysRevB.45.2207}
  {\bibfield  {journal} {\bibinfo  {journal} {Physical Review B}\ }\textbf
  {\bibinfo {volume} {45}},\ \bibinfo {pages} {2207} (\bibinfo {year}
  {1992})}\BibitemShut {NoStop}%
\bibitem [{\citenamefont {Waintal}(2006)}]{waintal_quantum_2006}%
  \BibitemOpen
  \bibfield  {author} {\bibinfo {author} {\bibfnamefont {X.}~\bibnamefont
  {Waintal}},\ }\bibfield  {title} {\bibinfo {title} {On the quantum melting of
  the two-dimensional {Wigner} crystal},\ }\href
  {https://doi.org/10.1103/PhysRevB.73.075417} {\bibfield  {journal} {\bibinfo
  {journal} {Physical Review B}\ }\textbf {\bibinfo {volume} {73}},\ \bibinfo
  {pages} {075417} (\bibinfo {year} {2006})}\BibitemShut {NoStop}%
\bibitem [{\citenamefont {Mora}\ and\ \citenamefont
  {Waintal}(2007)}]{mora_variational_2007}%
  \BibitemOpen
  \bibfield  {author} {\bibinfo {author} {\bibfnamefont {C.}~\bibnamefont
  {Mora}}\ and\ \bibinfo {author} {\bibfnamefont {X.}~\bibnamefont {Waintal}},\
  }\bibfield  {title} {\bibinfo {title} {Variational wave functions, ground
  state and their overlap},\ }\href
  {https://doi.org/10.1103/PhysRevLett.99.030403} {\bibfield  {journal}
  {\bibinfo  {journal} {Physical Review Letters}\ }\textbf {\bibinfo {volume}
  {99}},\ \bibinfo {pages} {030403} (\bibinfo {year} {2007})}\BibitemShut
  {NoStop}%
\bibitem [{\citenamefont {Nomura}(2021)}]{nomura_helping_2021}%
  \BibitemOpen
  \bibfield  {author} {\bibinfo {author} {\bibfnamefont {Y.}~\bibnamefont
  {Nomura}},\ }\bibfield  {title} {{\selectlanguage {english}\bibinfo {title}
  {Helping restricted {Boltzmann} machines with quantum-state representation by
  restoring symmetry}},\ }\href {https://doi.org/10.1088/1361-648X/abe268}
  {\bibfield  {journal} {\bibinfo  {journal} {J. Phys.: Condens. Matter}\
  }\textbf {\bibinfo {volume} {33}},\ \bibinfo {pages} {174003} (\bibinfo
  {year} {2021})}\BibitemShut {NoStop}%
\bibitem [{\citenamefont {Shi}\ \emph {et~al.}(2006)\citenamefont {Shi},
  \citenamefont {Duan},\ and\ \citenamefont {Vidal}}]{shi_classical_2006}%
  \BibitemOpen
  \bibfield  {author} {\bibinfo {author} {\bibfnamefont {Y.-Y.}\ \bibnamefont
  {Shi}}, \bibinfo {author} {\bibfnamefont {L.-M.}\ \bibnamefont {Duan}},\ and\
  \bibinfo {author} {\bibfnamefont {G.}~\bibnamefont {Vidal}},\ }\bibfield
  {title} {\bibinfo {title} {Classical simulation of quantum many-body systems
  with a tree tensor network},\ }\href
  {https://doi.org/10.1103/PhysRevA.74.022320} {\bibfield  {journal} {\bibinfo
  {journal} {Phys. Rev. A}\ }\textbf {\bibinfo {volume} {74}},\ \bibinfo
  {pages} {022320} (\bibinfo {year} {2006})}\BibitemShut {NoStop}%
\bibitem [{\citenamefont {Cheng}\ \emph {et~al.}(2019)\citenamefont {Cheng},
  \citenamefont {Wang}, \citenamefont {Xiang},\ and\ \citenamefont
  {Zhang}}]{cheng_tree_2019}%
  \BibitemOpen
  \bibfield  {author} {\bibinfo {author} {\bibfnamefont {S.}~\bibnamefont
  {Cheng}}, \bibinfo {author} {\bibfnamefont {L.}~\bibnamefont {Wang}},
  \bibinfo {author} {\bibfnamefont {T.}~\bibnamefont {Xiang}},\ and\ \bibinfo
  {author} {\bibfnamefont {P.}~\bibnamefont {Zhang}},\ }\bibfield  {title}
  {\bibinfo {title} {Tree tensor networks for generative modeling},\ }\href
  {https://doi.org/10.1103/PhysRevB.99.155131} {\bibfield  {journal} {\bibinfo
  {journal} {Physical Review B}\ }\textbf {\bibinfo {volume} {99}},\ \bibinfo
  {pages} {155131} (\bibinfo {year} {2019})}\BibitemShut {NoStop}%
\end{thebibliography}%

\end{document}